\documentclass[a4paper,11pt]{article}
\usepackage[left=2cm,right=2cm,top=2cm,bottom=2cm]{geometry}
\usepackage{psfrag,amsmath,amsfonts,amssymb,latexsym,amsthm,verbatim,color,lscape}
\usepackage[figuresright]{rotating}
\usepackage{moreverb}

\usepackage{multicol}
\usepackage{amsmath,lscape,amsfonts,amssymb,
amsthm,verbatim,color,lscape,psfrag}

\usepackage{url}
\usepackage{hyperref}
\hypersetup{colorlinks,
citecolor=black,
filecolor=black,
linkcolor=black,
urlcolor=black,
pdftex}

\usepackage{amsmath}
\usepackage{algorithm}
\usepackage[noend]{algpseudocode}

\usepackage[sort&compress,numbers]{natbib}

\usepackage{booktabs}

\usepackage{colortbl}
\definecolor{gray}{rgb}{0.8,0.8,0.8}

\usepackage{times}
\usepackage{blindtext}
\usepackage[font={footnotesize,it}]{caption}

\usepackage{sectsty}  
\subsectionfont{\normalsize\bf}
\sectionfont{\large\bf}

\def \red#1{\textcolor{red}{#1}}

\def \red#1{\textcolor{red}{#1}}

\begin{document}

\def\p{\partial}
\def\oo{\infty}
\def\rt#1{\sqrt{#1}\,}

\def\Cbar{{\overline C}}
\def\C{\mathbf{C}}
\def\E{{\rm E}\,}
\def\I{\mathbf{I}}
\def\pp{\mathbf{p}}
\def\R{\mathbf{R}}
\def\y{\mathbf{y}}
\def\Y{\mathbf{Y}}
\def\z{\mathbf{z}}
\def\x{\mathbf{x}}
\def\o{\omega}
\def\s{\sigma}

\def\V{\mathbf{V}}
\def\I{\mathbf{I}}
\def\bfv{\mathbf{v}}
\def\X{\mathbf{X}}
\def\D{\mathbf{D}}

\def\a{{\alpha}}
\def\g{\gamma}
\def\b{\beta}

\def\de{\delta}
\def\debf{\boldsymbol{\delta}}
\def\e{\epsilon}
\def\th{\theta}
\def\r{\rho}
\def\thbf{\boldsymbol{\theta}}
\def\taubf{\boldsymbol{\tau}}
\def\pibf{\boldsymbol{\pi}}
\def\Xibf{\boldsymbol{\Xi}}
\def\Sbf{\boldsymbol{\Sigma}}

\def \red#1{\textcolor{red}{#1}}
\def \blue#1{\textcolor{blue}{#1}}
\def \magenta#1{\textcolor{magenta}{#1}}
\def \green#1{\textcolor{green}{#1}}
\def\bbf{\boldsymbol{\beta}}
\def\bmu{\boldsymbol{\mu}}

\title{A multinomial quadrivariate D-vine copula mixed model for  meta-analysis of diagnostic studies in the presence of \\non-evaluable subjects}

\date{}
\author{
Aristidis K. Nikoloulopoulos\footnote{{\small\texttt{A.Nikoloulopoulos@uea.ac.uk}}, School of Computing Sciences, University of East Anglia, Norwich NR4 7TJ, UK} }
\maketitle

\begin{abstract}
\baselineskip=18pt
\noindent
Diagnostic test accuracy studies  observe the result of a gold standard procedure that defines the presence or absence of a disease and the result of a diagnostic test. They typically report the number of true positives, false positives, true negatives and false negatives. However, diagnostic test outcomes can also be either non-evaluable positives or non-evaluable negatives.  We propose a novel model  for meta-analysis of diagnostic studies in the presence of non-evaluable outcomes that assumes independent multinomial distributions for the true and non-evaluable positives,  and,   the true and non evaluable negatives, conditional on the latent sensitivity, specificity, probability of non-evaluable positives and  probability of non-evaluable negatives in each study. For the random effects distribution of the latent proportions, we employ a drawable vine copula that can  successively model the dependence in the joint tails. Our methodology is demonstrated with an extensive simulation study and applied  to data from diagnostic accuracy studies of coronary computed tomography  angiography for the detection of coronary artery disease. The comparison of our method with the existing approaches  yields findings in the real data application that change the current conclusions. 

\end{abstract}

\baselineskip=18pt

\section{Introduction}

Diagnostic test accuracy studies observe the result of a gold standard procedure that defines the presence or absence of a disease and the result of a diagnostic test. They typically report the number of true positives (diseased subjects correctly diagnosed), false positives (non-diseased subjects incorrectly diagnosed as diseased), true negatives (non-diseased subjects correctly diagnosed as non-diseased) and false negatives (diseased subjects incorrectly diagnosed as non-diseased). However, diagnostic test outcomes can be non-evaluable \citep{Begg-etal-1986}. This is the case for   coronary computed tomography (CT) angiography studies which have non-evaluable results of index text in various ways such as when transferring a segment/vessel  to a patient based evaluation \citep{Schuetz-etal-2012}.

Synthesis of diagnostic test accuracy studies is the most common medical application of multivariate meta-analysis \citep{JacksonRileyWhite2011,
MavridisSalanti13}. The purpose of a meta-analysis of diagnostic test accuracy studies is to combine information over different studies, and provide an integrated analysis that will have more statistical power to detect an accurate diagnostic test than an analysis based on a single study.
Nevertheless, 
the existence of  non-evaluable subjects  is an important issue that could  lead to biased meta-analytic estimates of index test accuracy \citep{Schuetz-etal-2012,ma-etal-2014, Nikoloulopoulos-2018-3dmeta-NE}.
Schuetz et al. \cite{Schuetz-etal-2012} studied different ad-hoc approaches dealing with diagnostic test non-evaluable subjects, such as non-evaluable subjects are excluded from the study, non-evaluable positives (non-evaluable diseased subjects)  are taken as true positives and non-evaluable negatives (non-evaluable non-diseased subjects) are taken as false positives, non-evaluable positives are taken as false negatives and non-evaluable negatives are taken as true negatives, and non-evaluable positives as false negatives and non-evaluable negatives as false positives.  In all of these approaches, Schuetz et al.  \cite{Schuetz-etal-2012} used the   bivariate generalized linear mixed model (BGLMM) \cite{Chu&Cole2006} and  concluded that excluding the index test non-evaluable subjects  leads to overestimation of  the meta-analytic estimates of sensitivity and specificity and recommended the  intent-to-diagnose approach by treating non-evaluable positives as false negatives and non-evaluable negatives as false positives.

Ma et al. \cite{ma-etal-2014}  used a trivariate generalized mixed model (TGLMM) approach by treating the non-evaluable subjects as missing data under a missing at random assumption (MAR). 
Ma et al. \cite{ma-etal-2014}, with extensive simulation studies,    showed that the intent-to-diagnose approach \citep{Schuetz-etal-2012} under-estimates both meta-analytic estimates of sensitivity and specificity, while the  TGLMM approach under the MAR assumption 
gives nearly unbiased estimates of sensitivity, specificity and prevalence.

Nikoloulopoulos \cite{Nikoloulopoulos-2018-3dmeta-NE}, similarly  with Ma et al.
\cite{ma-etal-2014}, extended the vine copula mixed model  for trivariate meta-analysis of diagnostic test accuracy studies accounting for disease prevalence
\cite{Nikoloulopoulos2015c}  to additionally account for non-evaluable subjects. 
The extended trivariate vine copula mixed model  includes the extended  TGLMM   as a special case and can also model sensitivity, specificity, and prevalence on the original scale.    Nikoloulopoulos  \cite{Nikoloulopoulos-2018-3dmeta-NE} demonstrated  that the extended  TGLMM leads to biased meta-analytic estimates of sensitivity, specificity and prevalence when the univariate random effects are misspecified and 
that the extended vine copula mixed model gives nearly unbiased estimates of test accuracy indices and disease prevalence.

\hyphenation{Nik-ol-oul-op-oul-os}

A recurrent theme underlying the  methodologies of Ma et al. \cite{ma-etal-2014} and Nikoloulopoulos \cite{Nikoloulopoulos-2018-3dmeta-NE} is  the need to make the MAR assumption,  that cannot be verified based on the observed data.  Hence, it is natural to be concerned about robustness or sensitivity of inferences to departures from the MAR assumption.   
The within-study model assumes that the 
number  of true negatives,  false negatives, false positives,  true positives, non-evaluable negatives, and    non-evaluable positives
 are multinomially distributed given  the  latent (random) vector of  sensitivity, specificity, disease prevalence, probability of non-evaluable positives and probability of non-evaluable negatives. Under the MAR assumption the multinomial probability mass function (pmf) 
decomposes into a product of independent  binomial pmfs given the random effects. 
Hence, the within-study model actually  assumes that the 
number  of true negatives,   number of   true positives,  number of diseased subjects,   number of non-evaluable negatives, and   number of  non-evaluable positives
 are conditionally independent and   binomially distributed given the random effects.
 The triplet of latent sensitivity, specificity and prevalence  are  independent of the missing probabilities, hence the joint likelihood factors   into two components, one involving only the  sensitivity, specificity and disease prevalence, and the other involving only the probabilities of non-evaluable positives  and non-evaluable negatives. Therefore, the methodology of Chu et al. \cite{chu-etal-2009} or  Nikoloulopoulos \cite{Nikoloulopoulos2015c} is  applied to the first likelihood component to infer about the sensitivity, specificity and disease prevalence. Hence, the models  in Ma et al. \cite{ma-etal-2014} and Nikoloulopoulos \cite{Nikoloulopoulos-2018-3dmeta-NE} extend the  BGLMM \cite{Chu&Cole2006} and the bivariate vine copula mixed model  \citep{Nikoloulopoulos2015b}, respectively, to the trivariate case by adding the disease prevalence as a third outcome to indirectly account for the non-evaluable results.
One the one hand the number of  diseased  subjects are binomially distributed with probability of success the latent prevalence and a support that includes the number of non-evaluable positives and the number of non-evaluable negatives, but the true positives and true negatives are binomially distributed with probability of success the latent sensitivity and specificity, respectively, and a support that does not  include either the number of non-evaluable positives or the number of non-evaluable negatives  on the other, just like in the BGLMM \cite{Chu&Cole2006} and the bivariate vine copula mixed model \citep{Nikoloulopoulos2015b}. Note in passing that a special case of the bivariate copula mixed model is the BGLMM, that is, a copula mixed model composed of a bivariate normal (BVN) copula with normal margins.

In this paper, in order to remedy this situation of ignoring the non-evaluable subjects in the derivation  of the meta-analytic estimates of  sensitivity and specificity,  we   include the number of  non-evaluable positives and the number of non-evaluable negatives  as separate non-missing response categories. 
Interestingly, the proposed model extends the  bivariate copula mixed model \citep{Nikoloulopoulos2015b} to the quadrivariate case  by directly adding the number of non-evaluable positives and number of non-evaluable negatives as a third and fourth outcome respectively. Hence, it directly utilizes all the available data. 
The bivariate copula mixed model  \cite{Nikoloulopoulos2015b} assumes independent binomial distributions for the true positives and true negatives, conditional on the latent pair of  sensitivity and specificity in each study. In the proposed methodology for the meta-analysis of diagnostic tests where we additionally account for non-evaluable outcomes of the diagnostic test, we will assume independent multinomial distributions for the true and non-evaluable positives,  and,   the true  and non evaluable negatives, conditional on the latent sensitivity, specificity, probability of non-evaluable positives and  probability of non-evaluable negatives in each study.

For the random effects distribution, we employ a  regular vine copula \citep{Bedford&Cooke02}. Regular vine copulas are suitable for high-dimensional data, 
hence given the low dimension $d=4$, where $d$ is the dimension we use their boundary case namely   a drawable vine (D-vine) copula. D-vine copulas  have become important in  many applications areas such as finance \cite{aasetal09,nikoloulopoulos&joe&li11}
and biological sciences \cite{Killiches&Czado-2018,nikoloulopoulos-2018-smmr}, to just name a few,  in order to deal with dependence in the joint tails. Another boundary case of regular vine copulas is the canonical vine copula, but this parametric family of copulas is only suitable if there exists a  (pilot) variable that drives the dependence among the variables  \cite{nikoloulopoulos&joe12,czado-etal-12-statMod},  which apparently is not the case in this application area.

The remainder of the paper proceeds as follows. 
Section \ref{model} introduces the multinomial quadrivariate D-vine copula mixed model for meta-analysis of diagnostic studies accounting for non-evaluable results, and provides computational details for maximum likelihood (ML) estimation.  Section \ref{simulations-section} studies the small-sample efficiency and robustness of the ML estimation of the   multinomial quadrivariate D-vine copula  mixed model.  Section \ref{application} applies our methodology to data from a  published meta-analysis for diagnostic accuracy studies of coronary computed tomography  angiography for the detection of coronary artery disease. We conclude with some discussion in Section \ref{discussion}, followed by a brief section with  software details.

\section{\label{model}The multinomial quadrivariate  D-vine copula mixed model}

In this section, we  introduce the  multinomial quadrivariate D-vine copula mixed model.  In Subsections \ref{norm-model} and \ref{beta-model}, a D-vine copula representation of the random effects distribution with normal and beta margins, respectively, is presented. 
We complete this section with details on maximum likelihood estimation.

\subsection{\label{norm-model}The multinomial  quadrivariate D-vine copula mixed model with normal margins}
 We first introduce the notation used in this paper. The data are  $y_{ijk},\, i = 1, . . . ,N,\, j=0,1,2,\,k=0,1$, where $i$ is an
index for the individual studies, $j$ is an index for the test outcome (0:negative; 1:positive;  2: non-evaluable) and $k$  is an index for the disease outcome (0: non-diseased; 1: diseased). 
The ``classic" $2\times 2$ table  is extended to a $3\times 2$ table (Table \ref{3times2}). 
Each cell in  Table \ref{3times2} provides  the cell frequency 
corresponding to a combination of index test and disease outcomes in study $i$.

\setlength{\tabcolsep}{35pt}
\begin{table}[!h]
\caption{\label{3times2} Data  from an individual study in a  $3\times 2$  table. }
\centering
\begin{tabular}{ll|cc|c}
\toprule
&&\multicolumn{2}{c|}{Disease (by gold standard)}\\ 
{Test}&  & $-$ & $+$&Total\\\hline
$-$&  & $y_{i00}$ &$y_{i01}$ & $y_{i0+}$ \\
$+$ & & $y_{i10}$&  $y_{i11}$ &$y_{i1+}$  \\
Non-evaluable  &&$y_{i20}$& $y_{i21}$ & $y_{i2+}$\\
\hline
Total &  &$y_{i+0}$&$y_{i+1}$&$y_{i++}$\\
\bottomrule
\end{tabular}

\end{table}

The diseased subjects have three possible states: false negative, true positive, and  non evaluable positive. The multinomial observation is therefore the number of diseased subjects   where the diagnostic test is in each of its states.  
Hence, we assume that  the  false negatives  $Y_{i01}$, the   true positives $Y_{i11}$,  and the non-evaluable positives  $Y_{i21}$ are 
multinomially distributed given $(X_1=x_1,X_3=x_3)$, viz.

\begin{multline}\label{within-diseased}
(Y_{i01},Y_{i11},Y_{i21})|(X_1=x_1,X_3=x_3)\sim\\ \mathcal{M}_3\Bigl(y_{i+1},1-l^{-1}(x_1,x_3)-l^{-1}(x_3,x_1),l^{-1}(x_1,x_3),l^{-1}(x_3,x_1)\Bigr),\footnotemark
\end{multline}
\footnotetext{$\mathcal{M}_T(n,$ $p_1, \dots,p_{T})$  is shorthand notation for the  multinomial distribution; $T$ is the number of cells, $n$ is the number of observations, and  $(p_1,\dots,p_T)$ with $p_1+\ldots+p_T = 1$ is the $T$-dimensional vector of success probabilities.}

\noindent where  $(X_1,X_3)$ is the  bivariate latent pair of transformed  sensitivity and  probability of non-evaluable positives and $l^{-1}(x_{j},x_{k})=\frac{e^{x_j}}{1+e^{x_j}+e^{x_k}}$ is the inverse multinomial logit link.

In a similar manner the non-diseased  subjects have also three possible states: true negative, and false positive, and  non-evaluable negative.  Hence we assume that 
the true negatives  $Y_{i00}$, the false positives $Y_{i10}$, and    the  non-evaluable negatives  $Y_{i20}$   are  multinomially distributed given $(X_2=x_2,X_4=x_4)$, viz.

\begin{multline}\label{within-nondiseased}
(Y_{i00},Y_{i10},Y_{i20})|(X_2=x_2,X_4=x_4)\sim \\\mathcal{M}_3\Bigl(y_{i+0},l^{-1}(x_2,x_4),1-l^{-1}(x_2,x_4)-l^{-1}(x_4,x_2),l^{-1}(x_4,x_2)\Bigr),
\end{multline}
where $(X_2,X_4)$ is the  bivariate latent pair of transformed specificity and  probability of non-evaluable negatives.

After defining the within-studies model in (\ref{within-diseased}) and (\ref{within-nondiseased}), we next define the between-studies model. 
The stochastic representation of the between studies model takes the form
\begin{equation}\label{copula-between-norm}
\Bigl(\Phi\bigl(X_1;l(\pi_1,\pi_3),\s_1^2\bigr),\Phi\bigl(X_2;l(\pi_2,\pi_4),\s_2^2\bigr),\Phi\bigl(X_3;l(\pi_3,\pi_1),\s_3^2\bigr),\Phi\bigl(X_4;l(\pi_4,\pi_2),\s_4^2\bigr)\Bigr)\sim C(\cdot;\thbf),
\end{equation}

\noindent where $C(\cdot;\thbf)$ is a quadrivariate  D-vine  copula with dependence parameter vector $\thbf=(\th_{12}, \th_{23},$ 
$\th_{34}, \th_{13|2}, \th_{24|3}, \th_{14|23})$ and $\Phi(\cdot;\mu,\s^2)$ is the cumulative distribution function (cdf) of the  N($\mu,\s^2$) distribution, and $l(\pi_j,\pi_k)=\log\Bigl(\frac{\pi_j}{1-\pi_j-\pi_k}\Bigr)$ is the multinomial logit link. The copula parameter vector $\thbf$ has parameters of the random effects model and they are separated from the univariate parameters $(\pi_j,\s_j),\,j=1,\ldots,4$. The parameters $\pi_1$ and $\pi_2$ are those of actual interest denoting the meta-analytic parameters for the sensitivity  and specificity, while the parameters $\pi_3$ and $\pi_4$ denote the probabilities of non-evaluable positives and negatives, respectively.  The univariate parameters $\s_1^2,\s^2_2, \s_3^2,\s^2_4$   denote the variabilities of the random effects. 

The
quadrivariate  D-vine copula is built via successive mixing from
 bivariate pair-copulas on different levels. The pairs at level 1 are $j,j+1$, for
$j=1,2,3$, and for level $\ell$ ($2\le\ell<4$), the (conditional)
pairs are $j,j+\ell|j+1,\ldots,j+\ell-1$ for $j=1,\ldots,4-\ell$.
That is, for the 4-dimensional D-vine, the copulas for variables $j$ and $j+\ell$ given the variables indexed in between capture the conditional dependence  \citep{nikoloulopoulos&joe&li11}.
When all the bivariate pair-copulas are BVN copulas with correlation (copula) parameters  $\rho_{12},\rho_{23},\rho_{34}$  (1st level) and partial correlation parameters $\rho_{13|2}, \rho_{24|3}, \rho_{14|23}$ (2nd and 3rd level),
the resulting distribution is the quadrivariate normal with mean vector $\bmu=\bigl(l(\pi_1,\pi_3),l(\pi_2,\pi_3),l(\pi_3,\pi_1),l(\pi_4,\pi_2)\bigr)^\top$ and variance covariance matrix
$\Sbf=\begin{pmatrix}
\sigma_1^2 &\rho_{12}\sigma_1\s_2 &\rho_{13}\sigma_1\s_3&\rho_{14}\sigma_1\s_4\\
\rho_{12}\sigma_1\sigma_2 & \sigma_2^2&\rho_{23}\sigma_2\s_3&\rho_{24}\sigma_2\s_4\\
\rho_{13}\sigma_1\sigma_3 &\rho_{23}\sigma_2\s_3& \sigma_3^2 &\rho_{34}\sigma_3\sigma_4\\
\rho_{14}\sigma_1\sigma_4&\rho_{24}\sigma_2\sigma_4  & \rho_{34}\sigma_3\sigma_4
&\sigma_4^2
\end{pmatrix},$ where
$\r_{13}=\r_{13|2}\sqrt{1-\r_{12}^2}\sqrt{1-\r_{23}^2} +\r_{12}\r_{23}$, $\r_{24}=\r_{24|3}\sqrt{1-\r_{23}^2}\sqrt{1-\r_{34}^2} + \r_{23}\r_{34}$,
$\r_{14}=\r_{14|2}\sqrt{1-\r_{12}^2}\sqrt{1-\r_{24}^2}+\r_{12}\r_{24}$, 
$\r_{14|2}=\r_{14|23}\sqrt{1-\r_{13|2}^2}\sqrt{1-\r_{34|2}^2}+\r_{13|2}\r_{34|2}$, 
$\r_{13|2}=(\r_{13}-\r_{12}\r_{23})/\sqrt{1-\r_{12}^2}/\sqrt{1-\r_{23}^2}$ and 
$\r_{34|2}=(\r_{34}-\r_{23}\r_{24})/\sqrt{1-\r_{23}^2}/\sqrt{1-\r_{24}^2}$.\cite{nikoloulopoulos-2018-smmr}
Other choices of copulas are better if there is more dependence   in joint upper or lower tail.

The models in (\ref{within-diseased}--\ref{copula-between-norm}) together specify a multinomial quadrivariate D-vine copula mixed  model with joint likelihood

\begin{align*}
&L(\pi_1,\pi_2,\pi_3,\pi_4,\s_1,\s_2,\s_3,\s_4,\thbf)
=\\
&\prod_{i=1}^N\int_{-\infty}^{\infty}\int_{-\infty}^{\infty}\int_{-\infty}^{\infty}\int_{-\infty}^{\infty}
g\Bigl(y_{i11},y_{i21};y_{i+1},l^{-1}(x_1,x_3),l^{-1}(x_3,x_1)\Bigr)\times\\
&g\Bigl(y_{i00},y_{i20};y_{i+0},l^{-1}(x_2,x_4),l^{-1}(x_4,x_2)\Bigr)f_{1234}(x_1,x_2,x_3,x_4;\thbf)dx_1dx_2dx_3dx_4, 
\end{align*}
where  $g(;n,p_1,\ldots,p_{T-1})$ is the $\mathcal{M}_T(n,p_1,\dots,p_{T})$ probability mass function (pmf) and  $f_{1234}(\cdot;\thbf)$ is the quadrivariate D-vine density, viz.
\begin{equation}\label{vine-density}
f_{1234}(x_1,x_2,x_3,x_4;\thbf)=
\phi(x_1)\phi(x_2)\phi(x_3)
  \phi(x_4)c_{1234}\bigl(\Phi(x_1),\Phi(x_2),\Phi(x_3),\Phi(x_4);\thbf\bigr),
\end{equation}
with 
\begin{align*}
&c_{1234}\bigl(\Phi(x_1),\Phi(x_2),\Phi(x_3),\Phi(x_4);\thbf\bigr)=\\
&c_{12}\bigl(\Phi(x_1),\Phi(x_2);\th_{12}\bigr)\,c_{23}\bigl(\Phi(x_2),\Phi(x_3);\th_{23}\bigr)
c_{34}\bigl(\Phi(x_3),\Phi(x_4);\th_{34}\bigr)\times\\  
& c_{13|2}\bigl(F_{1|2}(x_1|x_2),F_{3|2}(x_3|x_2);\th_{13|2}\bigr)c_{24|3}\bigl(F_{2|3}(x_2|x_3),F_{4|3}(x_4|x_3);\th_{24|3}\bigr)\times\\
&c_{14|23}\bigl(F_{1|23}(x_1|x_2,x_3),F_{4|23}(x_4|x_2,x_3);\th_{14|23}\bigr), 
\end{align*}
where $\phi(x)$ and $\Phi(x)$ is shorthand notation for  the density $\phi(x;\mu,\s^2)$ and cdf  $\Phi(x;\mu,\s^2)$ of the $N(\mu,\s^2)$ distribution, $c_{jk},c_{jk|\ell},c_{14|23}$ are bivariate copula densities, $F_{j|k}(x_j|x_k)=\frac{\partial C_{jk}\bigl(\Phi_j(x_j),\Phi_k(x_k)\bigr)}{\partial \Phi_k(x_k)}$, $F_{1|23}(x_1|x_2,x_3)=\frac{\partial C_{13|2}\bigl(F_{1|2}(x_1|x_2),F_{3|2}(x_3|x_2)\bigr)}{\partial \Phi(x_2)}$ and 
$F_{4|23}(x_4|x_2,x_3)=\frac{\partial C_{24|3}\bigl(F_{2|3}(x_2|x_3),F_{4|3}(x_4|x_3)\bigr) }{\partial \Phi(x_3)}$; $C_{jk},C_{jk|\ell}$ are bivariate copula cdfs. 
Note that a for a 4-dimensional D-vine copula density there are  12 different  decompositions \cite{aasetal09}. 
To be concrete in the exposition of the theory, we use the  decomposition in (\ref{vine-density}); the theory though also apply to the other 11 decompositions.

 Below we transform the original integral into an integral over a unit hypercube using the inversion method. Hence the joint likelihood becomes

\begin{align*}
&\prod_{i=1}^N\int_{0}^{1}\int_{0}^{1}\int_{0}^{1}\int_{0}^{1}
g\Bigl(y_{i11},y_{i21};y_{i+1},l^{-1}\bigl(\Phi^{-1}(u_1,l(\pi_1,\pi_3),\s_1^2),\Phi^{-1}(u_3,l(\pi_3,\pi_1),\s_3^2)\bigr),\\&l^{-1}\bigl(\Phi^{-1}(u_3,l(\pi_3,\pi_1),\s_3^2),\Phi^{-1}(u_1,l(\pi_1,\pi_3),\s_1^2)\bigr)\Bigr)g\Bigl(y_{i00},y_{i20};y_{i+0},l^{-1}\bigl(\Phi^{-1}(u_2,l(\pi_2,\pi_4),\s_2^2),\\&\Phi^{-1}(u_4,l(\pi_4,\pi_2),\s_4^2)\bigr),
l^{-1}\bigl(\Phi^{-1}(u_4,l(\pi_4,\pi_2),\s_4^2),\Phi^{-1}(u_2,l(\pi_2,\pi_4),\s_2^2)\bigr)\Bigr)c_{1234}(u_1,u_2,u_3,u_4;\thbf)d\mathbf{u}.
\end{align*}

\subsection{\label{beta-model}The multinomial D-vine copula mixed model with beta margins}

In this section we use the parametrization proposed by Wilson \cite{wilson2018} in order the latent sensitivity and specificity to remain on the original scale. 
The within-study model takes the form 

\begin{eqnarray}\label{withinBinom}
(Y_{i01},Y_{i11},Y_{i21}|(X_{1}=x_1,X_{3}=x_3)&\sim& \mathcal{M}_3\Bigl(y_{i+1},1-x_1-x_3(1-x_1),x_1,x_3(1-x_1)\Bigr);\nonumber\\
(Y_{i00},Y_{i10},Y_{i20}|(X_{2}=x_2,X_{4}=x_4)&\sim& 
\mathcal{M}_3\Bigl(y_{i+0},x_2,1-x_2-x_4(1-x_2),x_4(1-x_2)\Bigr).
\end{eqnarray}

The stochastic representation of the between studies model is
\begin{equation}\label{copula-between}
\Bigl(F(X_1;\pi_1,\g_1),F(X_2;\pi_2,\g_2),F(X_3;\frac{\pi_3}{1-\pi_1},\g_3),F(X_4;\frac{\pi_4}{1-\pi_2},\g_4)\Bigr)\sim C(\cdot;\thbf),
\end{equation}
where $C(\cdot;\thbf)$ is a  D-vine copula with dependence parameter vector $\thbf$ and $F(\cdot;\pi,\g)$ is the cdf of the Beta($\pi,\g$) distribution with $\pi$ the mean and $\g$ the dispersion parameter. The copula parameter vector $\thbf$ has the dependence parameters of the random effects model and they are separated from the univariate parameters $(\pi_j,\g_j),\,j=1,\ldots,4$. The parameters $\pi_1$ and $\pi_2$ are those of actual interest denoting the meta-analytic parameters for the sensitivity  and specificity, while the parameters $\pi_3$ and $\pi_4$ denote the probabilities of non-evaluable positives and negatives, respectively.  The univariate parameters $\g_1,\g_2, \g_3,\g_4$   denote the variabilities of the random effects.  In contrast with the model in the preceding subsection the random effects of sensitivity  and specificity are on the original scale.

The models in (\ref{withinBinom}) and (\ref{copula-between}) together specify a vine copula mixed  model with joint likelihood
\begin{align*}
&L(\pi_1,\pi_2,\pi_3,\pi_4,\g_1,\g_2,\g_3,\g_4,\thbf)
=\\
&\prod_{i=1}^N\int_{0}^{1}\int_{0}^{1}\int_{0}^{1}\int_{0}^{1}
g\Bigl(y_{i11},y_{i21};y_{i+1},x_1,x_3(1-x_1)\Bigr)g\Bigl(y_{i00},y_{i02};y_{i+0},x_2,x_4(1-x_2)\Bigr)\times\\&
f_{1234}(x_1,x_2,x_3,x_4;\thbf)dx_1dx_2dx_3dx_4,
\end{align*}
where $f_{1234}(\cdot;\thbf)$ is as in (\ref{vine-density}) where we use beta instead of normal marginal distributions.  Below we transform the integral into an integral over a unit hypercube using the inversion method. Hence the joint likelihood becomes

\begin{align*}
&\prod_{i=1}^N\int_{0}^{1}\int_{0}^{1}\int_{0}^{1}\int_{0}^{1}
g\Bigl(y_{i11},y_{i21};y_{i+1},F^{-1}(u_1;\pi_1,\g_1),F^{-1}(u_3;\frac{\pi_3}{1-\pi_1},\g_3)\bigr(1-F^{-1}(u_1;\pi_1,\g_1)\bigl)\Bigr)\times\\&
g\Bigl(y_{i00},y_{i20};y_{i+0},F^{-1}(u_2;\pi_2,\g_2),F^{-1}(u_4;\frac{\pi_4}{1-\pi_2},\g_4)\bigl(1-F^{-1}(u_2;\pi_2,\g_2)\bigr)\Bigr)\times\\&
c_{1234}(u_1,u_2,u_3,u_4;\thbf)du_1du_2du_3du_4.
\end{align*}

\subsection{\label{computation}Maximum likelihood estimation and computational details}

Estimation of the model parameters    can be approached by the standard maximum likelihood (ML) method, by maximizing the logarithm of the joint likelihood. 
The estimated parameters can be obtained by 
using a quasi-Newton \citep{nash90} method applied to the logarithm of the joint likelihood.  
This numerical  method requires only the objective
function, i.e.,  the logarithm of the joint likelihood, while the gradients
are computed numerically and the Hessian matrix of the second
order derivatives is updated in each iteration. The standard errors (SE) of the ML estimates can be also obtained via the gradients and the Hessian computed numerically during the maximization process.

For the multinomial quadrivariate D-vine copula mixed model numerical evaluation of the joint pmf can be achieved with the following steps:

\begin{enumerate}
\itemsep=10pt
\item Calculate Gauss-Legendre \citep{Stroud&Secrest1966}  quadrature points $\{u_q: q=1,\ldots,n_q\}$ 
and weights $\{w_q: q=1,\ldots,n_q\}$ in terms of standard uniform.
\item Convert from independent uniform random variables $\{u_{q_1}: q_1=1,\ldots,n_q\}$,  $\{u_{q_2}: q_2=1,\ldots,n_q\}$, $\{u_{q_3}: q_3=1,\ldots,n_q\}$, and $\{u_{q_4}: q_4=1,\ldots,n_q\}$ to   dependent uniform random variables $v_{q_1},v_{q_2|q_1},v_{q_3|q_1;q_2}$, and $v_{q_4|q_1;q_2,q_3}$ that have a D-vine distribution $C(\cdot;\thbf)$ using the algorithm in \cite{nikoloulopoulos-2018-smmr}:
\begin{algorithmic}[1]
\State Set  $v_{q_1}=u_{q_1}$
\State $v_{q_2|q_1}=C^{-1}_{2|1}(u_{q_2}|u_{q_1};\th_{12})$
\State $t_1=C_{1|2}(v_{q_1}|v_{q_2|q_1};\th_{12})$
\State $t_2=C_{3|1;2}^{-1}\left(u_{q_3}|t_1;\th_{12}),\th_{13|2}\right)$
\State $v_{q_3|q_1;q_2}=C_{3|2}^{-1}(t_2|v_{q_2|q_1};\th_{23})$
\State $t_3=C_{2|3}(v_{q_2|q_1}|v_{q_3|q_1;q_2};\th_{23})$
\State $t_4=C_{1|3;2}(t_1|t_2;\th_{13|2})$
\State $t_5=C_{4|1;2,3}(u_{q_4}|t_4;\th_{14|23})$
\State $t_6=C^{-1}_{4|2;3}(t_5|t_3;\th_{24|3})$
\State $v_{q_4|q_1;q_2,q_3}=C^{-1}_{4|3}(t_6|v_{q_3|q_1;q_2};\th_{34})$,
\end{algorithmic}
where  $C(v|u;\th)$ and  $C^{-1}(v|u;\th)$ are  conditional copula cdfs 
and their inverses.

\item Numerically evaluate the joint pmf, e.g.,

\begin{align*}
&\prod_{i=1}^N\int_{0}^{1}\int_{0}^{1}\int_{0}^{1}\int_{0}^{1}
g\Bigl(y_{i11},y_{i21};y_{i+1},F^{-1}(u_1;\pi_1,\g_1),F^{-1}(u_3;\frac{\pi_3}{1-\pi_1},\g_3)\bigr(1-F^{-1}(u_1;\pi_1,\g_1)\bigl)\Bigr)\times\\&
g\Bigl(y_{i00},y_{i20};y_{i+0},F^{-1}(u_2;\pi_2,\g_2),F^{-1}(u_4;\frac{\pi_4}{1-\pi_2},\g_4)\bigl(1-F^{-1}(u_2;\pi_2,\g_2)\bigr)\Bigr)\times\\&
c_{1234}(u_1,u_2,u_3,u_4;\thbf)du_1du_2du_3du_4. 
\end{align*}

\noindent in a quadruple sum:

\begin{align*}
&\sum_{q_1=1}^{n_q}\sum_{q_2=1}^{n_q}\sum_{q_3=1}^{n_q}
\sum_{q_4=1}^{n_q}w_{q_1}w_{q_2}w_{q_3}w_{q_4}
g\Bigl(y_{i11},y_{i21};y_{i+1},F^{-1}(v_{q_1};\pi_1,\g_1),\\
& F^{-1}(v_{q_3|q_1;q_2};\frac{\pi_3}{1-\pi_1},\g_3)\bigr(1-F^{-1}(v_{q_1};\pi_1,\g_1)\bigl)\Bigr)
g\Bigl(y_{i00},y_{i20};y_{i+0},
\\&F^{-1}(v_{q_2|q_1};\pi_2,\g_2),F^{-1}(v_{q_4|q_1;q_2,q_3};
\frac{\pi_4}{1-\pi_2},\g_4)\bigl(1-F^{-1}(v_{q_2|q_1};\pi_2,\g_2)\bigr)\Bigr). 
\end{align*}

\end{enumerate}

With Gauss-Legendre quadrature, the same nodes and weights
are used for different functions;
this helps in yielding smooth numerical derivatives for numerical optimization via quasi-Newton.

\section{\label{simulations-section}Simulations}
In this section we study the small-sample efficiency and robustness of the ML estimation of the   multinomial quadrivariate D-vine copula  mixed model. In Section \ref{simulations-1} we gauge the small-sample efficiency of the ML 
method, and  investigate 
the  misspecification of the parametric margin or  bivariate pair-copulas of the random effects distribution. 
In Section \ref{simulations-2} we  investigate the mixed model misspecification by using both the proposed model  
 and the extended  trivariate vine copula mixed model \cite{Nikoloulopoulos-2018-3dmeta-NE} as true models.

We set the sample size and the  true univariate  and dependence parameters to mimic the data analyzed in Section \ref{application}.  In each model, we use 6 different linking copula families: normal, Frank, and Clayton copula along with its rotated versions  (see our previous papers on copula mixed models
     \cite{Nikoloulopoulos2015b,Nikoloulopoulos2015c,Nikoloulopoulos-2016-SMMR,Nikoloulopoulos2018-AStA} 
  for definitions) to cover different types of dependence structure.
To make it easier to compare strengths of dependence, we convert the BVN, Frank and rotated Clayton estimated parameters to Kendall's $\tau$'s in $(-1, 1)$ via  the following relations  
\begin{equation*}\label{tauBVN}
\tau=\frac{2}{\pi}\arcsin(\th),
\end{equation*}
\begin{equation*}\label{tauFrank}
\tau=\left\{\begin{array}{ccc}
1-4\theta^{-1}-4\theta^{-2}\int_\theta^0\frac{t}{e^t-1}dt &,& \th<0\\
1-4\theta^{-1}+4\theta^{-2}\int^\theta_0\frac{t}{e^t-1}dt &,& \th>0\\
\end{array}\right., 
\end{equation*}
in \cite{HultLindskog02,genest87} and  
\begin{equation*}\label{tauCln}
\tau=\left\{\begin{array}{rcl}
\th/(\th+2)&,& \mbox{by 0$^\circ$ or 180$^\circ$ }\\
-\th/(\th+2)&,& \mbox{by 90$^\circ$ or 270$^\circ$}\\
\end{array}\right.. 
\end{equation*}
in \cite{genest&mackay86}, respectively. 

\subsection{\label{simulations-1}Small-sample efficiency--misspecification of the random effects distribution}

We randomly generate samples of size $N =30$  from the multinomial quadrivariate D-vine copula mixed model with  both  normal and beta margins. The simulation process is as below:

\begin{enumerate}
\itemsep=20pt
 \item Simulate $(u_1,u_2,u_3,u_4)$ from a D-vine distribution $C(\cdot;\tau_{12},\tau_{23},\tau_{34},\tau_{13|2}=0,\tau_{24|3}=0,\tau_{14|23}=0)$.

\item 

\begin{itemize}
\itemsep=20pt
\item Convert to  normal realizations via 
\begin{eqnarray*}
x_1=\Phi^{-1}\bigl(u_1;\log\frac{\pi_1}{1-\pi_1-\pi_3},\s_1\bigr) && x_2=\Phi^{-1}\bigl(u_2;\log\frac{\pi_2}{1-\pi_2-\pi_4},\s_2\bigr)\\
x_3=\Phi^{-1}\bigl(u_3;\log\frac{\pi_3}{1-\pi_1-\pi_3},\s_3\bigr) &&x_4=\Phi^{-1}\bigl(u_4;\log\frac{\pi_4}{1-\pi_2-\pi_4},\s_4\bigr). 
\end{eqnarray*}  
\item Convert to  beta realizations via 
\begin{eqnarray*}
x_1=F^{-1}\bigl(u_1;\pi_1,\g_1\bigr) && x_2=F^{-1}\bigl(u_2;\pi_2,\g_2\bigr)\\
x_3=F^{-1}\bigl(u_3;\frac{\pi_3}{1-\pi_1},\s_1\bigr)&& x_4=F^{-1}\bigl(u_4;\log\frac{\pi_4}{1-\pi_2},\g_4\bigr).
\end{eqnarray*}
\end{itemize} 
\item Simulate the  size of diseased and non-diseased subjects  $n_1$  and $n_2$, respectively, from a shifted gamma distribution  to obtain heterogeneous study sizes \citep{paul-etal-2010},  i.e., 
\begin{eqnarray*}
n_1&\sim& \mbox{sGamma}(\a=1.2,\b=0.01,\mbox{lag}=30)\\
n_2&\sim& \mbox{sGamma}(\a=1.2,\b=0.01,\mbox{lag}=30)
\end{eqnarray*}
and round off  $n_1$  and $n_2$ to the nearest integers. 
\item 
\begin{itemize}
\itemsep=20pt
\item For normal margins draw $(y_{01},y_{11},y_{21})$ from
$$\mathcal{M}_3\Bigl(n_1,\frac{1}{1+e^{x_1}+e^{x_3}},\frac{e^{x_1}}{1+e^{x_1}+e^{x_3}},\frac{e^{x_3}}{1+e^{x_1}+e^{x_3}}\Bigr)$$ and  $(y_{00},y_{10},y_{20})$ from 
$$ \mathcal{M}_3\Bigl(n_2,\frac{e^{x_2}}{1+e^{x_2}+e^{x_4}},\frac{1}{1+e^{x_2}+e^{x_4}},\frac{e^{x_4}}{1+e^{x_2}+e^{x_4}}\Bigr).$$

\item For beta margins draw $(y_{01},y_{11},y_{21})$ from  $$\mathcal{M}_3\Bigl(n_1,1-x_1-x_3(1-x_1),x_1,x_3(1-x_1)\Bigr)$$
 and  $(y_{00},y_{10},y_{20})$ from $$\mathcal{M}_3\Bigl(n_2,x_2,1-x_2-x_4(1-x_2),x_4(1-x_2)\Bigr).$$ 
\end{itemize}

\end{enumerate}

Tables  \ref{sim-beta} and \ref{sim-norm}   contain the 
resultant biases, root mean square errors (RMSE), and standard deviations (SD),  along with the square root of the average theoretical variances ($\sqrt{\bar V}$), scaled by 100, for the ML estimates  under different pair-copulas and marginal choices from  the multinomial D-vine copula mixed model with beta and normal margins, respectively. The true (simulated) pair-copula distributions are the Clayton  copulas  rotated by $180^\circ$  for both the $C_{12}(;\tau_{12})$ and $C_{34}(;\tau_{34})$ pair-copulas and the Clayton  copula  rotated by $90^\circ$  for  the $C_{23}(;\tau_{23})$ pair-copula. 

Conclusions from the values in the tables are the following:

\begin{itemize}
\itemsep=15pt
\item ML   with  the true multinomial D-vine copula mixed  model is highly efficient according to the simulated biases, SDs and RMSEs.

\item The ML estimates of the univariate meta-analytic parameters and their SDs are robust under copula misspecification, but are not robust to margin misspecification.

\item The ML estimates of $\tau$'s and their SDs are robust to copula misspecification, but they are not robust to margin misspecification.
 \end{itemize}

\begin{landscape}
\begin{table}[h!]

\setlength{\tabcolsep}{14pt}
  \centering
  \caption{\label{sim-beta}Small sample of sizes $N = 30$ simulations ($10^3$ replications; $n_q=15$) from the multinomial quadrivariate D-vine copula mixed model with  beta margins and resultant biases,  root mean square errors (RMSE) and standard deviations (SD), along with the square root of the average theoretical variances ($\sqrt{\bar V}$), scaled by 100,  for the ML estimates  under different pair-copula choices and margins.}

\begin{footnotesize}

    \begin{tabular}{lllccccccccccc}
    \toprule
          & \multicolumn{1}{l}{margin } & \multicolumn{1}{l}{copula} & $\pi_1=$ & $\pi_2=$ & $\pi_3=$ & $\pi_4=$ & $\g_1=$ & $\g_2=$ & $\g_3=$ & $\g_4=$ & $\tau_{12}=$ & $\tau_{23}=$ & $\tau_{34}=$ \\
                 &       &       & 0.90  & 0.77  & 0.06  & 0.11  & 0.09  & 0.08  & 0.37  & 0.15  & 0.82  & -0.52 & 0.26 \\
    \midrule
\rowcolor{gray}     Bias  &   normal & BVN   & 4.20 & 3.49 & -1.97 & -1.91 & -     & -     & -     & -     &-22.37	&36.27&16.07\\
          & beta  &       & -0.08 & -0.03 & 0.38 & 0.03 & -0.10 & -0.21 & -4.81 & -0.12 & -5.01 & 6.21 & 1.97 \\
\rowcolor{gray}           & normal & Frank & 4.24 & 3.68 & -1.96 & -1.86 & -     & -     & -     & -     & -21.28&34.84&15.74\\
          & beta  &       & 0.21 & 0.43 & 0.11 & -0.18 & -0.01 & -0.17 & -4.25 & -0.09 & -2.58 & 5.14 & 2.00 \\
   \rowcolor{gray}        & normal & Cln\{180$^\circ$,90$^\circ$\} & 4.20 & 3.37 & -2.00 & -1.84 & -     & -     & -     & -     & -15.90&30.22&	15.57\\
         & $^\dag$ beta  &       & -0.21 & -0.16 & 0.31 & 0.11 & -0.17 & -0.28 & -1.75 & -0.52 & 0.60 & 0.71 & 1.37 \\
 \rowcolor{gray}          & normal & Cln\{0$^\circ$,270$^\circ$\} & 4.14 & 3.52 & -1.90 & -1.85 & -     & -     & -     & -     &-30.10&	38.76&	12.30\\
          & beta  &       & -0.08 & 0.11 & 0.53 & 0.02 & 0.82 & 0.49 & -6.33 & 0.14 & -3.62 & 15.15 & -2.62 \\
\midrule   
\rowcolor{gray}  SD    & normal & BVN   & 1.84 & 2.68 & 1.59 & 1.74 & 24.50 & 14.06 & 28.18 & 17.58 & 24.52 & 27.58 & 25.93 \\

          & beta  &       & 1.95 & 2.53 & 1.71 & 1.67 & 2.97 & 2.28 & 8.29 & 4.35 & 10.26 & 14.27 & 17.16 \\
  \rowcolor{gray}         & normal & Frank & 1.89 & 2.74 & 1.65 & 1.81 & 24.70 & 14.04 & 28.44 & 17.80 & 24.90 & 28.57 & 26.45 \\
          & beta  &       & 1.84 & 2.37 & 1.61 & 1.58 & 3.00 & 2.22 & 8.53 & 4.34 & 8.02 & 14.71 & 17.31 \\
 \rowcolor{gray}          & normal & Cln\{180$^\circ$,90$^\circ$\} & 1.88 & 2.67 & 1.62 & 1.73 & 23.97 & 13.53 & 27.78 & 17.86 & 23.67 & 25.46 & 21.90 \\
       &  $^\dag$  beta  &       & 1.98 & 2.52 & 1.68 & 1.67 & 2.85 & 2.15 & 8.89 & 4.28 & 9.18 & 14.65 & 15.85 \\
 \rowcolor{gray}          & normal & Cln\{0$^\circ$,270$^\circ$\} & 1.88 & 2.76 & 1.62 & 1.79 & 26.28 & 15.89 & 30.75 & 18.60 & 33.78 & 28.34 & 30.21 \\
          & beta  &       & 1.98 & 2.63 & 1.74 & 1.71 & 3.59 & 2.83 & 9.05 & 4.54 & 16.13 & 16.23 & 19.53 \\
\midrule  \rowcolor{gray}   $\sqrt{\bar V}$  &    normal & BVN   & 1.38 & 2.39 & 1.17 & 1.66 & 16.86 & 10.85 & 25.40 & 14.73 & 15.55 & 15.62 & 15.66 \\
          & beta  &       & 1.34 & 1.99 & 1.21 & 1.46 & 1.97 & 1.82 & 7.92 & 4.06 & 9.04 & 13.14 & 14.88 \\
   \rowcolor{gray}        & normal & Frank & 1.31 & 2.28 & 1.12 & 1.62 & 16.21 & 10.76 & 24.94 & 14.66 & 13.13 & 13.75 & 14.50 \\
          & beta  &       & 1.18 & 1.85 & 1.10 & 1.36 & 1.84 & 1.94 & 8.25 & 4.05 & 7.84 & 13.07 & 15.20 \\
 \rowcolor{gray}          & normal & Cln\{180$^\circ$,90$^\circ$\} & 1.36 & 2.34 & 1.15 & 1.63 & 16.49 & 10.37 & 24.44 & 14.14 & 13.51 & 15.53 & 13.77 \\
         & $^\dag$ beta  &       & 1.33 & 1.99 & 1.21 & 1.44 & 1.92 & 1.83 & 8.00 & 3.97 & 8.08 & 13.45 & 14.33 \\
  \rowcolor{gray}         & normal & Cln\{0$^\circ$,270$^\circ$\} & 1.38 & 2.40 & 1.18 & 1.66 & 16.04 & 10.92 & 27.34 & 14.84 & 13.47 & 12.44 & 16.15 \\
          & beta  &       & 1.22 & 1.85 & 1.10 & 1.36 & 2.10 & 1.94 & 7.84 & 4.14 & 10.83 & 12.75 & 16.73 \\
\midrule  \rowcolor{gray}    RMSE  &  normal & BVN   & 4.59 & 4.40 & 2.53 & 2.58 & -     & -     & -     & -     &33.19&	45.56&	30.51 \\
          & beta  &       & 1.95 & 2.53 & 1.75 & 1.67 & 2.97 & 2.28 & 9.58 & 4.35 & 11.42 & 15.56 & 17.28 \\
   \rowcolor{gray}        & normal & Frank & 4.64 & 4.59 & 2.57 & 2.59 & -     & -     & -     & -     & 32.75&45.05&	30.78\\
          & beta  &       & 1.85 & 2.41 & 1.61 & 1.59 & 3.00 & 2.23 & 9.53 & 4.35 & 8.43 & 15.58 & 17.42 \\
  \rowcolor{gray}         & normal & Cln\{180$^\circ$,90$^\circ$\} & 4.60 & 4.30 & 2.58 & 2.53 & -     & -     & -     & -     &28.52&	39.52&	26.87 \\
         & $^\dag$ beta  &       & 1.99 & 2.52 & 1.70 & 1.67 & 2.85 & 2.17 & 9.06 & 4.31 & 9.20 & 14.67 & 15.91 \\
   \rowcolor{gray}        & normal & Cln\{0$^\circ$,270$^\circ$\} & 4.55 & 4.47 & 2.50 & 2.58 & -     & -     & -     & -     & 45.24&	48.01&	32.62 \\
          & beta  &       & 1.98 & 2.63 & 1.81 & 1.71 & 3.69 & 2.87 & 11.04 & 4.54 & 16.54 & 22.20 & 19.70 \\
    \bottomrule
    \end{tabular}%
  \end{footnotesize}
\vspace{-0.3cm}

\begin{flushleft}

\begin{footnotesize}
Cln\{$\omega_1^\circ,\omega_2^\circ$\}: The $C_{12}(\cdot;\tau_{12}),C_{34}(\cdot;\tau_{34})$ and $C_{23}(\cdot;\tau_{23})$ pair-copulas are Clayton rotated by $\omega_1$ and $\omega_2$ degrees, respectively;
$^\dag$: True model.
\end{footnotesize}  
\end{flushleft}  

\end{table}  
\end{landscape}

\begin{landscape}
\begin{table}[h!]

\setlength{\tabcolsep}{14pt}
  \centering  
  
  \caption{\label{sim-norm}Small sample of sizes $N = 30$ simulations ($10^3$ replications; $n_q=15$) from the multinomial quadrivariate D-vine copula mixed model with  normal margins and resultant biases,  root mean square errors (RMSE) and standard deviations (SD), along with the square root of the average theoretical variances ($\sqrt{\bar V}$), scaled by 100, for the ML estimates  under different pair-copula choices and margins.}

\begin{footnotesize}   
    \begin{tabular}{lllccccccccccc}
    \toprule
          & \multicolumn{1}{l}{margin } & \multicolumn{1}{l}{copula} & $\pi_1=$ & $\pi_2=$ & $\pi_3=$ & $\pi_4=$ & $\s_1=$ & $\s_2=$ & $\s_3=$ & $\s_4=$ & $\tau_{12}=$ & $\tau_{23}=$ & $\tau_{34}=$ \\
                    &       &       & 0.94  & 0.79  & 0.03  & 0.09  & 0.75  & 0.65  & 1.20  & 0.69  & 0.82  & -0.38 & 0.29 \\
    \midrule
  \rowcolor{gray}   Bias  & normal & BVN   & -0.64 & -0.33 & 0.61 & 0.25 & 0.99 & -1.22 & -5.03 & -0.88 & -6.98 & 4.30 & 5.50 \\
          & beta  &       & -6.16 & -4.21 & 4.08 & 2.29 & -     & -     & -     & -     &-15.26&	-13.26&	12.79 \\
 \rowcolor{gray}          & normal & Frank & -0.63 & -0.17 & 0.61 & 0.22 & 0.82 & -1.05 & -5.73 & -0.86 & -6.67 & 2.53 & 5.45 \\
          & beta  &       & -5.97 & -3.96 & 3.96 & 2.25 & -     & -     & -     & -     &-12.55&	-14.92&	12.18 \\
\rowcolor{gray}           & $^\dag$ normal & Cln\{180$^\circ$,90$^\circ$\}& -0.63 & -0.44 & 0.57 & 0.33 & -1.13 & -1.96 & -2.71 & -0.97 & -1.54 & -2.42 & 2.31 \\
          & beta  &       & -6.37 & -4.42 & 4.10 & 2.50 & -     & -     & -     & -     & -10.10&	-19.62&	9.71 \\
\rowcolor{gray}           & normal & Cln\{0$^\circ$,270$^\circ$\} & -0.72 & -0.24 & 0.71 & 0.24 & 3.57 & 1.36 & -3.63 & -0.46 & -4.08 & 11.78 & 4.52 \\
          & beta  &       & -6.20 & -4.25 & 4.23 & 2.37 & -     & -     & -     & -     &-21.47&	-6.61&	10.91\\
  \midrule \rowcolor{gray}   SD    & normal & BVN   & 2.12 & 2.75 & 1.83 & 1.84 & 18.29 & 11.62 & 23.06 & 14.40 & 17.54 & 17.42 & 19.35 \\
          & beta  &       & 2.99 & 2.94 & 2.31 & 1.94 & 5.26 & 3.00 & 6.51 & 3.42 & 10.98 & 18.63 & 22.53 \\
   \rowcolor{gray}        & normal & Frank & 2.20 & 2.80 & 1.91 & 1.88 & 17.92 & 11.60 & 23.42 & 14.50 & 14.46 & 18.54 & 20.16 \\
          & beta  &       & 2.97 & 3.00 & 2.35 & 2.00 & 5.23 & 3.17 & 6.71 & 3.43 & 10.68 & 19.30 & 22.22 \\
  \rowcolor{gray}         &$^\dag$ normal & Cln\{180$^\circ$,90$^\circ$\}& 2.14 & 2.77 & 1.84 & 1.86 & 17.74 & 11.44 & 22.79 & 14.36 & 15.47 & 19.08 & 16.82 \\
          & beta  &       & 3.06 & 3.03 & 2.34 & 2.01 & 5.16 & 3.25 & 7.07 & 3.50 & 11.46 & 20.35 & 21.01 \\
\rowcolor{gray}           & normal & Cln\{0$^\circ$,270$^\circ$\} & 2.15 & 2.81 & 1.85 & 1.86 & 19.92 & 13.08 & 24.72 & 15.13 & 22.16 & 19.47 & 24.34 \\
          & beta  &       & 2.99 & 3.02 & 2.33 & 1.98 & 5.73 & 3.39 & 6.60 & 3.46 & 16.13 & 21.83 & 30.25 \\
  \midrule \rowcolor{gray}   $\sqrt{\bar V}$  & normal & BVN   & 1.43 & 2.45 & 1.19 & 1.62 & 15.81 & 10.23 & 22.66 & 12.43 & 18.18 & 15.88 & 15.91 \\
          & beta  &       & 1.35 & 2.10 & 1.17 & 1.45 & 2.04 & 2.11 & 6.09 & 3.10 & 8.09 & 14.89 & 17.57 \\
     \rowcolor{gray}      & normal & Frank & 1.33 & 2.30 & 1.11 & 1.55 & 15.53 & 10.13 & 22.28 & 12.37 & 11.75 & 15.14 & 16.00 \\
          & beta  &       & 1.28 & 2.07 & 1.13 & 1.43 & 2.01 & 2.29 & 6.35 & 3.10 & 7.70 & 15.89 & 17.29 \\
   \rowcolor{gray}        &$^\dag$ normal & Cln\{180$^\circ$,90$^\circ$\} & 1.41 & 2.38 & 1.18 & 1.59 & 14.92 & 9.88 & 21.71 & 12.04 & 14.06 & 16.53 & 14.14 \\
          & beta  &       & 1.31 & 2.14 & 1.17 & 1.45 & 1.97 & 2.29 & 6.58 & 3.13 & 7.93 & 14.92 & 16.86 \\
   \rowcolor{gray}        & normal & Cln\{0$^\circ$,270$^\circ$\} & 1.39 & 2.41 & 1.17 & 1.60 & 16.20 & 10.56 & 23.22 & 12.61 & 18.50 & 15.09 & 18.85 \\
          & beta  &       & 1.26 & 1.95 & 1.08 & 1.34 & 2.20 & 2.09 & 5.72 & 3.15 & 8.89 & 18.63 & 20.45 \\
  \midrule \rowcolor{gray}   RMSE  & normal & BVN   & 2.22 & 2.77 & 1.93 & 1.86 & 18.32 & 11.68 & 23.60 & 14.42 & 18.88 & 17.94 & 20.12 \\
          & beta  &       & 6.85 & 5.13 & 4.69 & 3.00 & -     & -     & -     & -     &18.80&	22.87&	25.91 \\
    \rowcolor{gray}       & normal & Frank & 2.29 & 2.81 & 2.00 & 1.89 & 17.93 & 11.65 & 24.11 & 14.53 & 15.93 & 18.71 & 20.88 \\
          & beta  &       & 6.67 & 4.96 & 4.61 & 3.01 & -     & -     & -     & -     & 16.48&	24.39&	25.34 \\
  \rowcolor{gray}         &$^\dag$ normal & Cln\{180$^\circ$,90$^\circ$\} & 2.23 & 2.80 & 1.93 & 1.89 & 17.77 & 11.61 & 22.95 & 14.39 & 15.55 & 19.24 & 16.98 \\
          & beta  &       & 7.07 & 5.36 & 4.72 & 3.21 & -     & -     & -     & -     & 15.28&	28.27&	23.14 \\
 \rowcolor{gray}          & normal & Cln\{0$^\circ$,270$^\circ$\} & 2.27 & 2.82 & 1.98 & 1.88 & 20.24 & 13.15 & 24.98 & 15.13 & 22.53 & 22.75 & 24.76 \\
          & beta  &       & 6.88 & 5.22 & 4.83 & 3.09 & -     & -     & -     & -     & 26.85&	22.81&	32.15 \\
    \bottomrule
    \end{tabular}%
    \end{footnotesize}
\vspace{-0.3cm}     
    
    \begin{flushleft}

\begin{footnotesize}
Cln\{$\omega_1^\circ,\omega_2^\circ$\}: The $C_{12}(\cdot;\tau_{12}),C_{34}(\cdot;\tau_{34})$ and $C_{23}(\cdot;\tau_{23})$ pair-copulas are Clayton rotated by $\omega_1$ and $\omega_2$ degrees, respectively;
$^\dag$: True model. 
\end{footnotesize}  
\end{flushleft}  
  \end{table}  
\end{landscape}

\subsection{\label{simulations-2}Misspecification of the copula mixed model that accounts for non-evaluable outcomes}

\baselineskip=23pt

We randomly generate samples of size $N =30$  from the multinomial quadrivariate D-vine copula mixed model and the extended trivariate vine copula mixed model with  both  normal and beta margins  using the algorithm in Section \ref{simulations-1} and in Nikoloulopoulos \cite{Nikoloulopoulos-2018-3dmeta-NE}, respectively. 
We compare the ML estimates of common parameters for both approaches under misspecification and  also include in the comparison the bivariate copula mixed model estimates where  the non-evaluable positives and negatives  are either excluded or included as false negatives and false positives (intention to diagnose approach), respectively.

In Section \ref{simulations-1} and in  Nikoloulopoulos \cite{Nikoloulopoulos-2018-3dmeta-NE}, it has been revealed  that (a) the estimation of the univariate meta-analytic parameters is a univariate inference, and hence it is the univariate marginal distribution that matters and not the type of the copula, and (b)  estimated Kendall's $\tau$ is similar amongst different families of copulas. Hence as the ML estimates are nearly not  affected by the type of the pair-copula, we provide here the results when all the bivariate copulas are BVN.

Tables  \ref{comparison-beta} and \ref{comparison-norm}   contain the 
resultant biases, RMSEs, and SDs,  along with the square root of the average theoretical variances ($\sqrt{\bar V}$), scaled by 100, for the ML estimates  under different copula mixed models. The true quadrivariate multinomial vine copula mixed model is composed by the Clayton  copulas  rotated by $180^\circ$  for both the $C_{12}(;\tau_{12})$ and $C_{34}(;\tau_{34})$ pair-copulas and the Clayton  copula  rotated by $90^\circ$  for  the $C_{23}(;\tau_{23})$ pair-copula.   The true trivariate  vine copula mixed model is composed by the Clayton  copula for $C_{12}(;\tau_{12})$ and  the  Clayton rotated by $90^\circ$  for both the $C_{13}(;\tau_{13})$ and $C_{23|1}(;\tau_{23|1})$ pair-copulas.  

\bigskip

 Conclusions from the values in the tables are the following:
 
\begin{itemize}
\item The bivariate copula mixed model where the non-evaluable outcomes are disregarded and the extended trivariate vine  copula mixed model showed similar performance. Both led to unbiased (biased) and efficient (inefficient)  estimates when the  true model is the trivariate  (quadrivariate multinomial) vine copula mixed model.

\item The bivariate copula mixed model where the non-evaluable positives and negatives included as false negatives and false positives, respectively,  and the multinomial D-vine copula mixed model with beta margins showed similar performance. Both led to unbiased (biased) and efficient (inefficient)  estimates when the  true model is the quadrivariate multinomial  vine copula mixed model with beta margins (trivariate vine copula mixed model or quadrivariate multinomial  vine copula mixed model with normal margins).

\end{itemize}

\begin{landscape}
\begin{table}[htbp]
\setlength{\tabcolsep}{14pt}
  \centering
  \caption{\label{comparison-beta}Small sample of sizes $N = 30$ simulations ($10^3$ replications; $n_q=15$) from the multinomial quadrivariate D-vine and trivariate  vine copula mixed models with  beta margins and resultant biases,  root mean square errors (RMSE) and standard deviations (SD), along with the square root of the average theoretical variances ($\sqrt{\bar V}$), scaled by 100, for the ML estimates  under different copula mixed models.}
 
\begin{footnotesize}    
    \begin{tabular}{rrrcccccccccccc}
    \toprule
     {} & {} & {} &       & \multicolumn{11}{c}{True  vine copula mixed model}  \\
     \cmidrule{5-15}
    {} & {} & {} &       & \multicolumn{5}{c}{Trivariate } &       & \multicolumn{5}{c}{Quadrivariate } \\
  \cmidrule{5-9}
          \cmidrule{11-15}
    \multicolumn{1}{l}{} & \multicolumn{1}{l}{Fitted copula} & \multicolumn{1}{l}{margin} &       & $\pi_1=$ & $\pi_2=$ & $\g_1=$ & $\g_2=$ & $\tau=$ &       & $\pi_1=$ & $\pi_2=$ & $\g_1=$ & $\g_2=$ & $\tau=$ \\
    \multicolumn{1}{l}{} & \multicolumn{1}{l}{mixed model} & \multicolumn{1}{l}{} &       & 0.97  & 0.85  & 0.03  & 0.06  & 0.39  &       & 0.90  & 0.77  & 0.09  & 0.08  & 0.82 \\\hline
\rowcolor{gray}     \multicolumn{1}{l}{Bias} & \multicolumn{1}{l}{Bivariate$^\P$ } & \multicolumn{1}{l}{beta} &       & 0.04  & 0.22  & -0.11 & -0.15 & 11.37 &       & 7.10  & 9.63  & -5.84 & -2.25 & -42.16 \\
    \multicolumn{1}{l}{} & \multicolumn{1}{l}{} & \multicolumn{1}{l}{normal$^\dag$} &       & 0.91  & 2.38  & -     & -     & 14.24 &       & 8.26  & 11.78 & -     & -     & -40.39 \\
   \rowcolor{gray}  \multicolumn{1}{l}{} & \multicolumn{1}{l}{Bivariate$^\S$} & \multicolumn{1}{l}{beta} &       & -3.18 & -5.46 & 1.70  & 0.14  & -2.54 &       & -0.08 & -0.03 & -0.08 & -0.22 & -4.75 \\
    \multicolumn{1}{l}{} & \multicolumn{1}{l}{} & \multicolumn{1}{l}{normal$^\dag$} &       & -1.59 & -3.50 & -     & -     & -1.26 &       & 2.79  & 1.47  & -     & -     & -2.35 \\
 \rowcolor{gray}    \multicolumn{1}{l}{} & \multicolumn{1}{l}{Trivariate} & \multicolumn{1}{l}{beta} &       & -0.03 & -0.09 & -0.10 & -0.06 & 8.97  &       & 7.10  & 9.60  & -5.81 & -2.24 & -42.47 \\
    \multicolumn{1}{l}{} & \multicolumn{1}{l}{} & \multicolumn{1}{l}{normal$^{\ddag}$} &       & 0.86  & 2.13  & -     & -     & 10.89 &       & 8.25  & 11.76 & -     & -     & -40.61 \\
\rowcolor{gray}     \multicolumn{1}{l}{} & \multicolumn{1}{l}{Quadrivariate} & \multicolumn{1}{l}{beta} &       & -3.18 & -5.46 & 1.71  & 0.13  & -1.63 &       & -0.08 & -0.03 & -0.10 & -0.21 & -5.01 \\
    \multicolumn{1}{l}{} & \multicolumn{1}{l}{} & \multicolumn{1}{l}{normal} &       & -0.46 & -1.10 & -     & -     & 21.73 &       & 4.20 & 3.49 & -     & -     & -22.37 \\\hline
\rowcolor{gray}     \multicolumn{1}{l}{SD} & \multicolumn{1}{l}{Bivariate$^\P$} & \multicolumn{1}{l}{beta} &       & 0.64  & 1.91  & 1.29  & 2.00  & 24.08 &       & 0.81  & 1.83  & 1.91  & 1.99  & 17.07 \\
    \multicolumn{1}{l}{} & \multicolumn{1}{l}{} & \multicolumn{1}{l}{normal$^\dag$} &       & 0.57  & 1.90  & 23.49 & 14.40 & 25.41 &       & 0.49  & 1.72  & 23.15 & 13.93 & 17.40 \\
 \rowcolor{gray}   \multicolumn{1}{l}{} & \multicolumn{1}{l}{Bivariate$^\S$} & \multicolumn{1}{l}{beta} &        & 0.77  & 1.85  & 1.58  & 1.92  & 17.45 &       & 1.94  & 2.54  & 2.94  & 2.22  & 10.03 \\
    \multicolumn{1}{l}{} & \multicolumn{1}{l}{} & \multicolumn{1}{l}{normal$^\dag$} &       & 0.80  & 1.94  & 19.38 & 12.22 & 18.29 &       & 1.87  & 2.86  & 25.11 & 14.07 & 9.06 \\
\rowcolor{gray}     \multicolumn{1}{l}{} & \multicolumn{1}{l}{Trivariate} & \multicolumn{1}{l}{beta} &       & 0.66  & 1.89  & 1.27  & 2.02  & 22.71 &       & 0.81  & 1.83  & 1.93  & 1.99  & 16.92 \\
    \multicolumn{1}{l}{} & \multicolumn{1}{l}{} & \multicolumn{1}{l}{normal$^{\ddag}$} &       & 0.58  & 1.88  & 23.45 & 14.46 & 23.79 &       & 0.48  & 1.72  & 23.51 & 14.10 & 17.35 \\
\rowcolor{gray}     \multicolumn{1}{l}{} & \multicolumn{1}{l}{Quadrivariate} & \multicolumn{1}{l}{beta} &       & 0.77  & 1.87  & 1.57  & 1.93  & 18.20 &       & 1.95 & 2.53 & 2.97 & 2.28 & 10.25 \\
    \multicolumn{1}{l}{} & \multicolumn{1}{l}{} & \multicolumn{1}{l}{normal} &       & 0.68  & 1.91  & 23.35 & 13.83 & 23.92 &       & 1.84 & 2.68 & 24.50 & 14.06 & 24.90 \\\hline
\rowcolor{gray}     \multicolumn{1}{l}{$\sqrt{\bar V}$} & \multicolumn{1}{l}{Bivariate$^\P$} & \multicolumn{1}{l}{beta} &       & 0.63  & 1.80  & 1.34  & 2.02  & 27.90 &       & 0.60  & 1.61  & 1.21  & 1.69  & 15.20 \\
    \multicolumn{1}{l}{} & \multicolumn{1}{l}{} & \multicolumn{1}{l}{normal$^\dag$} &       & 0.53  & 1.73  & 24.78 & 14.05 & 26.16 &       & 0.45  & 1.52  & 19.40 & 11.91 & 15.35 \\
\rowcolor{gray}     \multicolumn{1}{l}{} & \multicolumn{1}{l}{Bivariate$^\S$} & \multicolumn{1}{l}{beta} &       & 1.08  & 2.09  & 1.78  & 2.00  & 16.39 &       & 1.31  & 1.97  & 1.93  & 1.79  & 8.79 \\
    \multicolumn{1}{l}{} & \multicolumn{1}{l}{} & \multicolumn{1}{l}{normal$^\dag$} &       & 0.93  & 2.10  & 19.55 & 11.99 & 17.09 &       & 1.27  & 2.04  & 16.09 & 10.47 & 8.03 \\
\rowcolor{gray}     \multicolumn{1}{l}{} & \multicolumn{1}{l}{Trivariate} & \multicolumn{1}{l}{beta} &       & 0.66  & 1.86  & 1.37  & 2.06  & 25.75 &       & 0.60  & 1.60  & 1.20  & 1.68  & 15.06 \\
    \multicolumn{1}{l}{} & \multicolumn{1}{l}{} & \multicolumn{1}{l}{normal$^\ddag$} &       & 0.54  & 1.77  & 24.03 & 14.24 & 22.36 &       & 0.45  & 1.52  & 19.13 & 11.82 & 15.12 \\
\rowcolor{gray}     \multicolumn{1}{l}{} & \multicolumn{1}{l}{Quadrivariate} & \multicolumn{1}{l}{beta} &       & 1.09  & 2.09  & 1.78  & 2.00  & 17.08 &       & 1.34 & 1.99 & 1.97 & 1.82 & 9.04 \\
    \multicolumn{1}{l}{} & \multicolumn{1}{l}{} & \multicolumn{1}{l}{normal} &       & 0.81  & 2.01  & 24.81 & 13.61 & 22.07 &       & 1.38 & 2.39 & 16.86 & 10.85 & 15.55 \\\hline
\rowcolor{gray}     \multicolumn{1}{l}{RMSE} & \multicolumn{1}{l}{Bivariate$^\P$} & \multicolumn{1}{l}{beta} &       & 0.64  & 1.92  & 1.29  & 2.00  & 26.63 &       & 7.15  & 9.80  & 6.14  & 3.00  & 45.48 \\
    \multicolumn{1}{l}{} & \multicolumn{1}{l}{} & \multicolumn{1}{l}{normal$^\dag$} &       & 1.08  & 3.04  & -     & -     & 29.13 &       & 8.28  & 11.91 & -     & -     & 43.98 \\
 \rowcolor{gray}    \multicolumn{1}{l}{} & \multicolumn{1}{l}{Bivariate$^\S$} & \multicolumn{1}{l}{beta} &       & 3.27  & 5.77  & 2.32  & 1.92  & 17.64 &       & 1.94  & 2.54  & 2.94  & 2.23  & 11.10 \\
    \multicolumn{1}{l}{} & \multicolumn{1}{l}{} & \multicolumn{1}{l}{normal$^\dag$} &       & 1.78  & 4.01  & -     & -     & 18.34 &       & 3.36  & 3.21  & -     & -     & 9.36 \\
\rowcolor{gray}     \multicolumn{1}{l}{} & \multicolumn{1}{l}{Trivariate} & \multicolumn{1}{l}{beta} &       & 0.67  & 1.89  & 1.28  & 2.02  & 24.42 &       & 7.15  & 9.77  & 6.12  & 3.00  & 45.72 \\
    \multicolumn{1}{l}{} & \multicolumn{1}{l}{} & \multicolumn{1}{l}{normal$^\ddag$} &       & 1.04  & 2.84  & -     & -     & 26.16 &       & 8.27  & 11.88 & -     & -     & 44.16 \\
\rowcolor{gray}    \multicolumn{1}{l}{} & \multicolumn{1}{l}{Quadrivariate} & \multicolumn{1}{l}{beta} &       & 3.27  & 5.77  & 2.32  & 1.94  & 18.27 &       & 1.95 & 2.53 & 2.97 & 2.28 & 11.42 \\
    \multicolumn{1}{l}{} & \multicolumn{1}{l}{} & \multicolumn{1}{l}{normal} &       & 0.82  & 2.21  & -     & -     & 32.32 &       & 4.59 & 4.40 & -     & -     & 33.19\\
    \bottomrule
    \end{tabular}%
    \end{footnotesize}  
    
    \vspace{-0.3cm}     
    
   \begin{flushleft}
   \begin{footnotesize}
$^\P$: The non-evaluable outcomes are excluded; $^\S$: The non-evaluable positives and negatives are included as false negatives and positives, respectively; $^\dag$: The resulting model is the same as the BGLMM; $^\ddag$: The resulting model is the same as the extended TGLMM. 
\end{footnotesize}  
\end{flushleft}  
\end{table}
\end{landscape}

\begin{landscape}
\begin{table}[htbp]
  \centering
  \caption{\label{comparison-norm}Small sample of sizes $N = 30$ simulations ($10^3$ replications; $n_q=15$) from the multinomial quadrivariate D-vine and trivariate  vine copula mixed models with  normal margins and resultant biases,  root mean square errors (RMSE) and standard deviations (SD), along with the square root of the average theoretical variances ($\sqrt{\bar V}$), scaled by 100, for the ML estimates  under different copula mixed models.}
  \setlength{\tabcolsep}{14pt}
  \begin{footnotesize}   
    \begin{tabular}{lllcccccccccccc}
    \toprule
     {} & {} & {} &       & \multicolumn{11}{c}{True  vine copula mixed model}  \\
     \cmidrule{5-15}
    {} & {} & {} &       & \multicolumn{5}{c}{Trivariate } &       & \multicolumn{5}{c}{Quadrivariate } \\
  \cmidrule{5-9}
          \cmidrule{11-15}
    \multicolumn{1}{l}{} & \multicolumn{1}{l}{Fitted copula} &    margin &       & $\pi_1=$ & $\pi_2=$ & $\s_1=$ & $\s_2=$ & $\tau=$ &       & $\pi_1=$ & $\pi_2=$ & $\s_1=$ & $\s_2=$ & $\tau=$ \\
          &           mixed model &  &       & 0.98  & 0.88  & 0.90  & 0.73  & 0.39  &       & 0.94  & 0.79  & 0.75  & 0.65  & 0.82 \\\hline
 \rowcolor{gray}    Bias  & Bivariate$^\P$  & beta  &       & -0.85 & -1.91 & -     & -     & 13.57 &       & 2.69  & 6.59  & -     & -     & -4.63 \\
          &       & normal$^\dag$ &       & -0.03 & 0.17  & -6.52 & -1.39 & 15.81 &       & 3.35  & 8.18  & -1.80 & -1.96 & -4.69 \\
 \rowcolor{gray}          & Bivariate$^\S$ & beta  &       & -4.04 & -7.58 & -     & -     & -1.43 &       & -5.97 & -3.98 & -     & -     & -17.33 \\
          &       & normal$^\dag$ &       & -2.46 & -5.61 & -0.79 & -6.12 & -0.35 &       & -1.75 & -1.93 & 63.85 & 17.11 & -14.39 \\
  \rowcolor{gray}         & Trivariate & beta  &       & -0.92 & -2.19 & -     & -     & 10.97 &       & 2.63  & 6.45  & -     & -     & -7.20 \\
          &       & normal$^\ddag$ &       & -0.08 & -0.07 & -6.28 & -1.15 & 12.26 &       & 3.31  & 8.09  & -1.64 & -2.09 & -7.31 \\
 \rowcolor{gray}          & Quadrivariate & beta  &       & -4.04 & -7.57 & -     & -     & -0.13 &       & -6.16 & -4.21 & -     & -     & -15.26 \\
          &       & normal &       & -1.39 & -3.28 & -8.23 & -3.69 & 23.31 &       & -0.64 & -0.33 & 0.99 & -1.22 & -6.98 \\\hline
 \rowcolor{gray}    SD    & Bivariate$^\P$ & beta  &       & 0.69  & 1.94  & 1.43  & 2.13  & 25.07 &       & 0.54  & 1.61  & 0.80  & 1.56  & 17.37 \\
          &       & normal$^\dag$ &       & 0.52  & 1.75  & 24.43 & 13.65 & 26.48 &       & 0.44  & 1.48  & 17.34 & 11.50 & 16.85 \\
\rowcolor{gray}           & Bivariate$^\S$ & beta  &       & 0.78  & 1.83  & 1.54  & 1.94  & 17.78 &       & 2.95  & 2.86  & 5.50  & 3.08  & 10.51 \\
          &       & normal$^\dag$ &       & 0.76  & 1.84  & 18.23 & 11.72 & 18.50 &       & 2.11  & 2.96  & 25.41 & 13.97 & 9.93 \\
 \rowcolor{gray}          & Trivariate & beta  &       & 0.72  & 1.92  & 1.43  & 2.15  & 23.29 &       & 0.60  & 1.69  & 0.86  & 1.55  & 17.14 \\
          &       & normal$^\ddag$ &       & 0.54  & 1.73  & 24.00 & 13.63 & 24.34 &       & 0.47  & 1.55  & 17.87 & 11.40 & 16.62 \\
\rowcolor{gray}           & Quadrivariate & beta  &       & 0.79  & 1.84  & 1.53  & 1.96  & 18.77 &       & 2.99 & 2.94 & 5.26 & 3.00 & 10.98 \\
          &       & normal &       & 0.66  & 1.78  & 24.11 & 13.34 & 25.91 &       & 2.12 & 2.75 & 18.29 & 11.62 & 17.54 \\\hline
\rowcolor{gray}     $\sqrt{\bar V}$ & Bivariate$^\P$  & beta  &       & 0.59  & 1.75  & 1.18  & 1.93  & 29.79 &       & 0.50  & 1.48  & 0.71  & 1.38  & 19.52 \\
          &       & normal$^\dag$ &       & 0.50  & 1.67  & 23.60 & 13.40 & 30.27 &       & 0.45  & 1.45  & 16.23 & 10.60 & 23.19 \\
  \rowcolor{gray}         & Bivariate$^\S$ & beta  &       & 1.06  & 2.06  & 1.71  & 1.97  & 16.68 &       & 1.33  & 2.10  & 2.01  & 2.13  & 8.30 \\
          &       & normal$^\dag$ &       & 0.91  & 2.07  & 18.92 & 11.82 & 16.94 &       & 1.34  & 2.22  & 15.15 & 11.02 & 8.07 \\
 \rowcolor{gray}          & Trivariate & beta  &       & 0.61  & 1.78  & 1.20  & 1.94  & 25.49 &       & 0.51  & 1.48  & 0.72  & 1.37  & 18.94 \\
          &       & normal$^\ddag$ &       & 0.51  & 1.70  & 22.86 & 13.25 & 22.97 &       & 0.45  & 1.45  & 15.63 & 10.37 & 19.41 \\
\rowcolor{gray}           & Quadrivariate & beta  &       & 1.06  & 2.07  & 1.70  & 1.98  & 16.83 &       & 1.35 & 2.10 & 2.04 & 2.11 & 8.09 \\
          &       & normal &       & 0.79  & 1.96  & 22.00 & 12.84 & 24.51 &       & 1.43 & 2.45 & 15.81 & 10.23 & 18.18 \\\hline
 \rowcolor{gray}    RMSE  & Bivariate$^\P$ & beta  &       & 1.10  & 2.72  & -     & -     & 28.51 &       & 2.74  & 6.79  & -     & -     & 17.98 \\
          &       & normal$^\dag$ &       & 0.52  & 1.75  & 25.28 & 13.72 & 30.84 &       & 3.38  & 8.31  & 17.43 & 11.66 & 17.49 \\
\rowcolor{gray}           & Bivariate$^\S$ & beta  &       & 4.11  & 7.80  & -     & -     & 17.83 &       & 6.66  & 4.90  & -     & -     & 20.27 \\
          &       & normal$^\dag$ &       & 2.58  & 5.91  & 18.25 & 13.22 & 18.50 &       & 2.74  & 3.53  & 68.73 & 22.09 & 17.49 \\
 \rowcolor{gray}          & Trivariate & beta  &       & 1.17  & 2.91  & -     & -     & 25.74 &       & 2.70  & 6.67  & -     & -     & 18.59 \\
          &       & normal$^\ddag$ &       & 0.55  & 1.73  & 24.81 & 13.67 & 27.25 &       & 3.35  & 8.23  & 17.94 & 11.59 & 18.16 \\
 \rowcolor{gray}          & Quadrivariate & beta  &       & 4.12  & 7.79  & -     & -     & 18.77 &       & 6.85 & 5.13 & -     & -     & 18.80 \\
          &       & normal &       & 1.54  & 3.73  & 25.47 & 13.84 & 34.85 &       & 2.22 & 2.77& 18.32& 11.68 & 18.88 \\
    \bottomrule
    \end{tabular}%
    \end{footnotesize}  
\vspace{-0.3cm}    
    
  \begin{flushleft}    
 \begin{footnotesize}
$^\P$: The non-evaluable outcomes are excluded; $^\S$: The non-evaluable positives and negatives are included as false negatives and positives, respectively; $^\dag$: The resulting model is the same as the BGLMM; $^\ddag$: The resulting model is the same as the extended TGLMM. 
\end{footnotesize}  
\end{flushleft}  
\end{table}
\end{landscape}

\baselineskip=18pt

\section{\label{application}Meta-analysis of coronary computed tomography angiography studies}

We apply the multinomial quadrivariate D-vine copula mixed model for the meta-analysis of diagnostic accuracy studies accounting for non-evaluable subjects  to data on 30 studies from a  systematic review for diagnostic accuracy studies of coronary computed tomography  angiography for the detection of coronary artery disease \citep{Menke&Kowalski2016}.  
 
We fit the multinomial quadrivariate D-vine copula mixed model for all different  decompositions of the D-vine copula density, for both beta and normal margins and different pair copulas at the level 1; for levels 2 and 3 we use BVN copulas. 
In cases when fitting the multinomial quadrivariate D-vine copula mixed model, the resultant estimate of one of the conditional dependence  parameters was close to the right boundary of its parameter space (that is clear indication that the model with a full structure provides more dependence structure than it is actually required  \cite{Nikoloulopoulos2015c}),   we used a truncated model, i.e.,  we captured the strongest dependence in the first tree and then just used the independence copulas in lower order trees, i.e. conditional independence. 
Joe et al. \cite{joeetal10} showed that in order for a vine copula to have (tail) dependence for all bivariate margins, it is only necessary for the bivariate copulas in level 1 to have (tail) dependence and it is not necessary for the conditional bivariate copulas in levels 2 and 3, to have tail dependence. Hence one can either use BVN or independence 
copulas at levels 2 and 3 without sacrificing the tail dependence of the vine copula distribution.

In Table \ref{app-table} we present the results from   the  decomposition of the vine copula density in (\ref{vine-density}), as a different decompositions led to similar results 
due to the small sample size. This is consistent with our previous studies on vine copula mixed models  \cite{Nikoloulopoulos2015c,Nikoloulopoulos-2018-3dmeta-NE}.
Since the number of parameters is not  the same between the models, we use the AIC, that  is $-2\times$log-likelihood $+2\times$ (\#model parameters)
as a rough diagnostic measure for goodness of fit between the models.
The AICs showed that a (truncated) multinomial quadrivariate D-vine copula mixed model with Clayton  copulas  rotated by $180^\circ$  for both the $C_{12}(;\tau_{12})$ and $C_{34}(;\tau_{34})$ pair-copulas and the Clayton  copula  rotated by $90^\circ$  for  the $C_{23}(;\tau_{23})$ pair-copula and beta margins (Table \ref{app-table})  provides the best fit.

\begin{table}[!h]
\setlength{\tabcolsep}{10.5pt}
  \centering
  \caption{\label{app-table}AICs, ML estimates and standard errors (SE) of the  multinomial quadrivariate  D-vine copula mixed models for  diagnostic accuracy studies of coronary computed tomography angiography.}
    \begin{tabular}{ccccccccccccc}
    \toprule
   \multicolumn{13}{l}{Normal margins} \\

          &       & \multicolumn{2}{c}{BVN} &       & \multicolumn{2}{c}{Frank} &       & \multicolumn{2}{c}{  Cln\{180$^\circ$,90$^\circ$\}} &       & \multicolumn{2}{c}{Cln\{180$^\circ$,270$^\circ$\}}\\
          &       & Est.  & SE    &       & Est.  & SE    &       & Est.  & SE    &       & Est.  & SE \\  \cmidrule{3-4}\cmidrule{6-7}\cmidrule{9-10}\cmidrule{12-13}
    $\pi_1$ &       & 0.94  & 0.01  &       & 0.95  & 0.01  &       & 0.94  & 0.02  &       & 0.94  & 0.02 \\
    $\pi_2$ &       & 0.80  & 0.03  &       & 0.80  & 0.03  &       & 0.79  & 0.03  &       & 0.79  & 0.03 \\
    $\pi_3$ &       & 0.04  & 0.01  &       & 0.03  & 0.01  &       & 0.03  & 0.01  &       & 0.04  & 0.01 \\
    $\pi_4$ &       & 0.09  & 0.02  &       & 0.09  & 0.02  &       & 0.09  & 0.02  &       & 0.09  & 0.02 \\
    $\s_1$ &       & 0.89  & 0.20  &       & 0.91  & 0.19  &       & 0.75  & 0.17  &       & 0.83  & 0.17 \\
    $\s_2$ &       & 0.72  & 0.15  &       & 0.65  & 0.13  &       & 0.65  & 0.12  &       & 0.67  & 0.13 \\
    $\s_3$ &       & 1.32  & 0.36  &       & 1.37  & 0.36  &       & 1.20  & 0.31  &       & 1.19  & 0.33 \\
    $\s_4$ &       & 0.80  & 0.23  &       & 0.70  & 0.21  &       & 0.69  & 0.19  &       & 0.73  & 0.19 \\
    $\tau_{12}$ &       & 0.54  & 0.22  &       & 0.49  & 0.20  &       & 0.82  & 0.19  &       & 0.82  & 0.18 \\
    $\tau_{23}$ &       & -0.16 & 0.20  &       & -0.31 & 0.17  &       & -0.38 & 0.24  &       & -0.04 & 0.15 \\
    $\tau_{34}$ &       & 0.22  & 0.23  &       & 0.11  & 0.24  &       & 0.29  & 0.17  &       & 0.37  & 0.17 \\
    $\tau_{13|2}$ &       & 0.43  & 0.34  &       & 0.67  & 0.23  &       & -     & -     &       & -     & - \\
    $\tau_{24|3}$ &       & 0.11  & 0.22  &       & -0.03 & 0.24  &       & -     & -     &       & -     & - \\
    $\tau_{14|23}$ &       & -0.39 & 0.32  &       & -0.36 & 0.49  &       & -     & -     &       & -     & - \\\midrule
    AIC   &       & \multicolumn{2}{c}{4013.22} &       & \multicolumn{2}{c}{4010.80} &       & \multicolumn{2}{c}{4007.72} &       & \multicolumn{2}{c}{4009.36} \\
    
 \midrule
    \multicolumn{13}{l}{Beta margins} \\
          &       & \multicolumn{2}{c}{BVN} &       & \multicolumn{2}{c}{Frank} &       & \multicolumn{2}{c}{ $^\dag$ Cln\{180$^\circ$,90$^\circ$\}} &       & \multicolumn{2}{c}{Cln\{180$^\circ$,270$^\circ$\}} \\
          &       & Est.  & SE    &       & Est.  & SE    &       & Est.  & SE    &       & Est.  & SE \\ \cmidrule{3-4}\cmidrule{6-7}\cmidrule{9-10}\cmidrule{12-13}
    $\pi_1$ &       & 0.90  & 0.02  &       & 0.90  & 0.02  &       & 0.90  & 0.01  &       & 0.89  & 0.01 \\
    $\pi_2$ &       & 0.76  & 0.03  &       & 0.77  & 0.02  &       & 0.77  & 0.02  &       & 0.76  & 0.02 \\
    $\pi_3$ &       & 0.06  & 0.01  &       & 0.06  & 0.01  &       & 0.06  & 0.01  &       & 0.07  & 0.01 \\
    $\pi_4$ &       & 0.11  & 0.02  &       & 0.11  & 0.02  &       & 0.11  & 0.02  &       & 0.11  & 0.02 \\
    $\g_1$ &       & 0.08  & 0.03  &       & 0.09  & 0.03  &       & 0.09  & 0.03  &       & 0.10  & 0.03 \\
    $\g_2$ &       & 0.09  & 0.03  &       & 0.09  & 0.02  &       & 0.08  & 0.02  &       & 0.09  & 0.02 \\
    $\g_3$ &       & 0.32  & 0.12  &       & 0.32  & 0.13  &       & 0.37  & 0.12  &       & 0.28  & 0.12 \\
    $\g_4$ &       & 0.15  & 0.07  &       & 0.16  & 0.07  &       & 0.15  & 0.07  &       & 0.15  & 0.06 \\
    $\tau_{12}$ &       & 0.71  & 0.11  &       & 0.74  & 0.08  &       & 0.82  & 0.08  &       & 0.79  & 0.07 \\
    $\tau_{23}$ &       & -0.35 & 0.17  &       & -0.34 & 0.12  &       & -0.52 & 0.14  &       & -0.23 & 0.10 \\
    $\tau_{34}$ &       & 0.23  & 0.22  &       & 0.20  & 0.21  &       & 0.26  & 0.18  &       & 0.21  & 0.17 \\
    $\tau_{13|2}$ &       & -0.66 & 0.38  &       & -     & -     &       & -     & -     &       & -     & - \\
    $\tau_{24|3}$ &       & -0.10 & 0.20  &       & -     & -     &       & -     & -     &       & -     & - \\
    $\tau_{14|23}$ &       & -0.02 & 0.57  &       & -     & -     &       & -     & -     &       & -     & - \\\midrule
    AIC   &       & \multicolumn{2}{c}{4009.42} &       & \multicolumn{2}{c}{4005.93} &       & \multicolumn{2}{c}{4002.17} &       & \multicolumn{2}{c}{4004.92} \\
    \bottomrule
    \end{tabular}%
    \hspace{-1cm}
     \begin{flushleft}
     
\begin{footnotesize}
Cln\{$\omega_1^\circ,\omega_2^\circ$\}: The $C_{12}(\cdot;\tau_{12}),C_{34}(\cdot;\tau_{34})$ and $C_{23}(\cdot;\tau_{23})$ pair copulas are Clayton rotated by $\omega_1$ and $\omega_2$ degrees, respectively;
$^\dag$: Best fit. 
\end{footnotesize}  
\end{flushleft}  
\end{table}

In real data (in contrast with the simulated data in Section \ref{simulations-section}), the truth is unknown, so it is important to compare between the proposed and other existing  approaches in terms of point estimation and variance. First, in order to reveal if the use of the proposed  model is worthy,   when a standard bivariate analysis (either ignoring the non-evaluable outcomes or including the non-evaluable positives and negatives  as false negatives and positives, respectively)   is  easy, we also fit the bivariate copula mixed model  \cite{Nikoloulopoulos2015b} with  both beta and normal margins and different bivariate copulas. 
According to the likelihood principle a bivariate copula mixed model with a  Clayton and Clayton copula rotated by 180$^\circ$  (to model  the association between the latent sensitivity and specificity)  and beta margins provides the best fit for both different ad-hoc approaches to handle the non-evaluable outcomes (Table \ref{other-analyses}). It is  revealed that a bivariate copula mixed model with the sensitivity and  specificity on the original scale provides better fit than the BGLMM \cite{Chu&Cole2006}, which models the sensitivity and  specificity on a transformed scale.

Then in order to compare the proposed approach with the  ones that use the MAR assumption 
we fit the extended trivariate vine copula mixed model  \cite{Nikoloulopoulos-2018-3dmeta-NE} with  both beta and normal margins and different pair copulas. 
According to the likelihood principle a vine copula mixed model composed of a Clayton copula  to model the association between the sensitivity and specificity,  a Clayton copula rotated by 90$^\circ$ to model both the associations between the specificity and prevalence and  between the sensitivity and  prevalence given the specificity, and beta margins provides the best fit (Table \ref{other-analyses}). It is revealed that an extended trivariate  vine copula mixed model with the sensitivity, specificity, and disease prevalence on the original scale provides better fit than the  extended TGLMM  \cite{ma-etal-2014}, which models the sensitivity, specificity, and disease prevalence on a transformed scale.

\begin{table}[htbp]
  \centering
  \caption{\label{other-analyses}AICs, ML estimates and standard errors (SE) of the best fitted bivariate copula  and extended trivariate vine copula mixed models with beta margins for  diagnostic accuracy studies of coronary computed tomography angiography.}
  \setlength{\tabcolsep}{17pt}
    \begin{tabular}{ccccrcccp{2cm}p{2cm}p{2cm}p{2cm}}
    \toprule
          &       & \multicolumn{5}{c}{Bivariate }        & &\multicolumn{4}{c}{Trivariate } \\
\cmidrule{1-1}    \cmidrule{3-7}\cmidrule{9-12}
          &       & \multicolumn{2}{c}{$^\P$ Clayton} & \multicolumn{1}{c}{} & \multicolumn{2}{c}{$^\S$ Clayton $180^\circ$} &       & \multicolumn{4}{c}{Clayton$\{0^\circ,90^\circ\}$} \\
          
          &       & Est.  & SE    & \multicolumn{1}{c}{} & Est.  & SE    &       & \multicolumn{2}{c}{Est.} & \multicolumn{2}{c}{SE} \\
\cmidrule{1-1}           \cmidrule{3-4}\cmidrule{6-7} \cmidrule{9-12}
    $\pi_1$ &       & 0.97  & 0.01  &       & 0.90  & 0.01  & \multicolumn{1}{c}{} & \multicolumn{2}{c}{0.97} & \multicolumn{2}{c}{0.01} \\
    $\pi_2$ &       & 0.85  & 0.02  &       & 0.77  & 0.02  & \multicolumn{1}{c}{} & \multicolumn{2}{c}{0.85} & \multicolumn{2}{c}{0.02} \\
    $\pi_3$ &       & -     & -     &       & -     & -     & \multicolumn{1}{c}{} & \multicolumn{2}{c}{0.49} & \multicolumn{2}{c}{0.03} \\
    $\g_1$ &       & 0.03  & 0.01  &       & 0.09  & 0.03  & \multicolumn{1}{c}{} & \multicolumn{2}{c}{0.03} & \multicolumn{2}{c}{0.01} \\
    $\g_2$ &       & 0.06  & 0.02  &       & 0.08  & 0.02  & \multicolumn{1}{c}{} & \multicolumn{2}{c}{0.06} & \multicolumn{2}{c}{0.02} \\
    $\g_3$ &       & -     & -     &       & -     & -     & \multicolumn{1}{c}{} & \multicolumn{2}{c}{0.11} & \multicolumn{2}{c}{0.02} \\
    $\tau_{12}$ &       & 0.42  & 0.19  &       & 0.82  & 0.08  & \multicolumn{1}{c}{} & \multicolumn{2}{c}{0.39} & \multicolumn{2}{c}{0.20} \\
    $\tau_{13}$ &       & -     & -     &       & -     & -     & \multicolumn{1}{c}{} & \multicolumn{2}{c}{0.02} & \multicolumn{2}{c}{0.23} \\
    $\tau_{23|1}$ &       & -     & -     &       & -     & -     & \multicolumn{1}{c}{} & \multicolumn{2}{c}{-0.28} & \multicolumn{2}{c}{0.17} \\\cmidrule{1-1} \cmidrule{3-4}\cmidrule{6-7} \cmidrule{9-12}
    AIC &       & \multicolumn{2}{c}{244.82} & \multicolumn{1}{c}{} & \multicolumn{2}{c}{321.91} &       & \multicolumn{4}{c}{492.26} \\
    \bottomrule
    \end{tabular}%
     \begin{flushleft}
     \begin{footnotesize}
    $^\P$: The non-evaluable outcomes are excluded; $^\S$: The non-evaluable positives and negatives are included as false negatives and positives, respectively.
    \end{footnotesize}
    \end{flushleft}
  \label{tab:addlabel}%
\end{table}

It has been shown that the trivariate analysis does not  change the conclusions from the bivariate analysis excluding the non-evaluable outcomes. It is also apparent that the results from the   quadrivariate analysis differentiate from the ones from bivariate (excluding the non-evaluable outcomes) and trivariate analysis which are fairly similar. The meta-analytic estimates of sensitivity and specificity from the latter approaches are blown, because in  both approaches it is assumed that 
$$
Y_{i11}|X_{1}=x_1\sim\mbox{Binomial}\bigl(y_{i01}+y_{i11},x_1\bigr) \quad \mbox{and}
\quad Y_{i00}|X_{2}=x_2\sim\mbox{Binomial}\bigl(y_{i00}+y_{i10},x_2\bigr),
$$
i.e., their support  ignores the number of non-evaluable positives $y_{i21}$ and the number of non-evaluable negatives  $y_{i20}$.  The conclusions from the quadrivariate  analysis with the latent proportions on the original scale  are quite similar with the ones from the  bivariate analysis where the the non-evaluable positives and negatives  are included as false negatives and positives, respectively. These results are consistent  with the findings in the simulations in Section \ref{simulations-2}.
 Note in passing that comparing the AIC values   amongst the qudrivariate, trivariate and bivariate copula mixed models is inconclusive as they use a different number of responses.

Though typically the focus of meta-analysis has been to derive the  summary-effect estimates, there is increasing interest in drawing predictive inference. 
Summary receiver operating characteristic curves
(SROC)  can be deduced from the D-vine copula mixed model with the sensitivity and specificity on the original scale  through the quantile regression techniques developed for the bivariate copula mixed model \cite{Nikoloulopoulos2015b}. SROC essentially shows the effect of different  model (random effect distribution) assumptions, since it is an inference that depends on the joint distribution.
An SROC curve has been deduced for the bivariate copula mixed model \cite{Nikoloulopoulos2015b} through a median regression  curve of $X_1$ on $X_2$. 
For the copula mixed model, the model parameters (including dependence parameters), the choice of the copula, and the choice of the margin affect the shape of the SROC curve \citep{Nikoloulopoulos2015b}. 
However, there is no priori reason to regress $X_1$ on $X_2$ instead of the other way around, so  a median regression  curve of $X_2$ on $X_1$ has also been provided. 
Rucker and Schumacher \cite{Rucker-schumacher-2009} stated that instead of summarizing data using an SROC, it might be preferable  to give confidence regions.
Hence, in addition to using just median regression curves, quantile regression curves with a focus on high ($q$ = 0.99) and low quantiles ($q$ = 0.01), which are strongly associated with the upper and lower tail dependence imposed from each parametric family of copulas, have been proposed  \cite{Nikoloulopoulos2015b}. These can been seen as confidence regions of the median regression SROC curve.

\begin{landscape}
\begin{figure}[!htbp]
\caption{\label{SROCs} SROC curves  with a confidence  region and summary operating points  (a pair of the model-based sensitivity  and specificity)  from the best fitted multinomial quadrivariate D-vine,  extended trivariate vine and bivariate copula mixed models  along with the study estimates. } 
\begin{center}
\begin{tabular}{cc}
\hline
Bivariate $^\P$ &  Trivariate \\\hline
\includegraphics[width=0.55\textwidth]{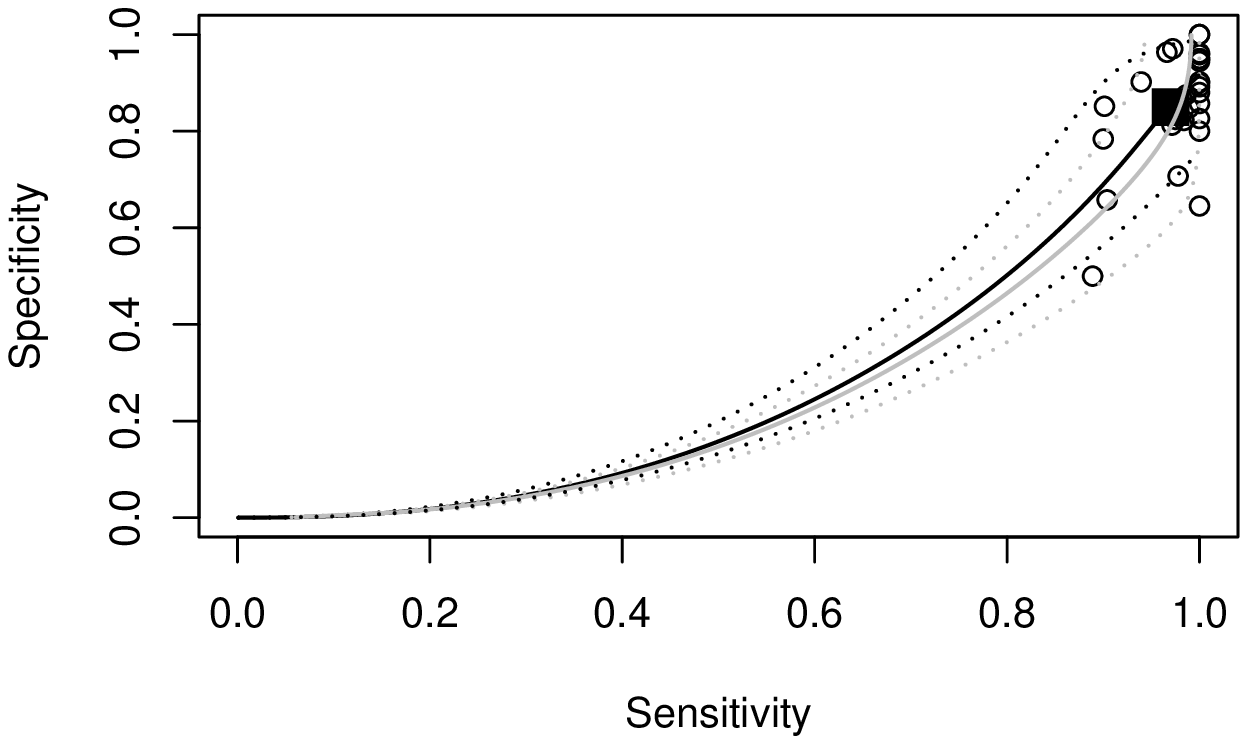}&
\includegraphics[width=0.55\textwidth]{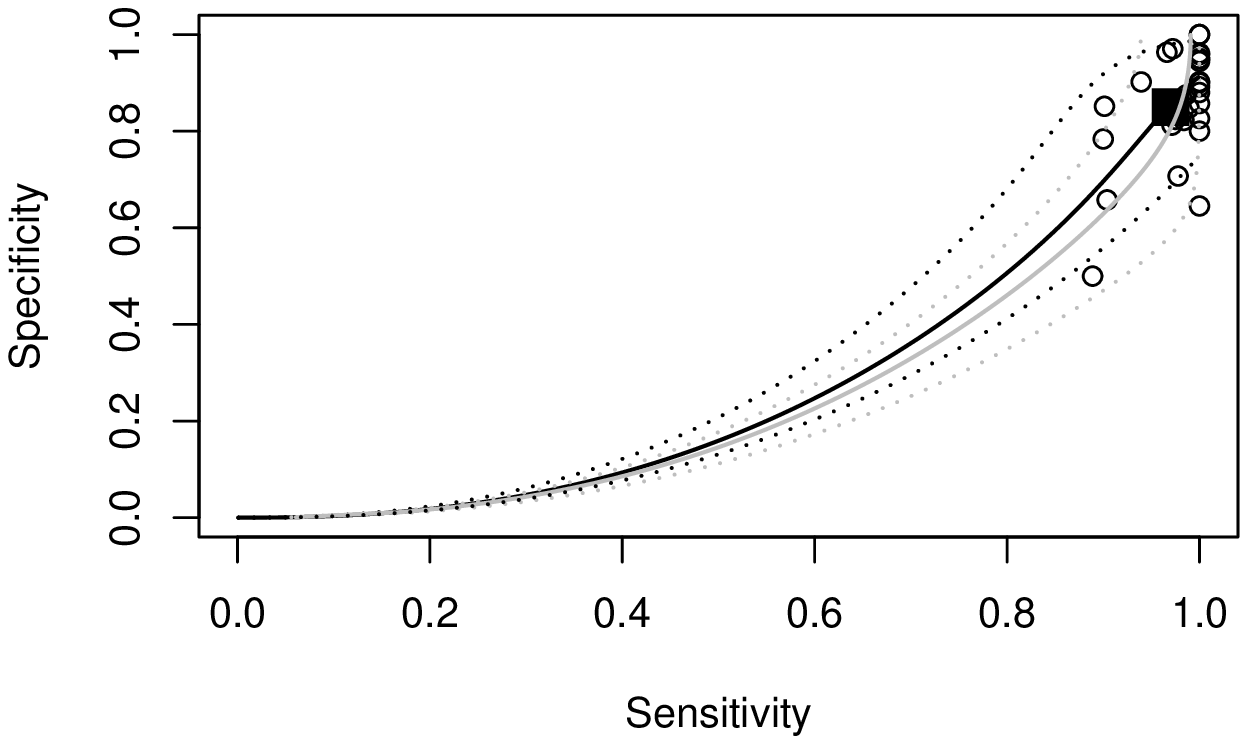} \\\hline
 Bivariate $^\S$ &Quadrivariate \\\hline

\includegraphics[width=0.55\textwidth]{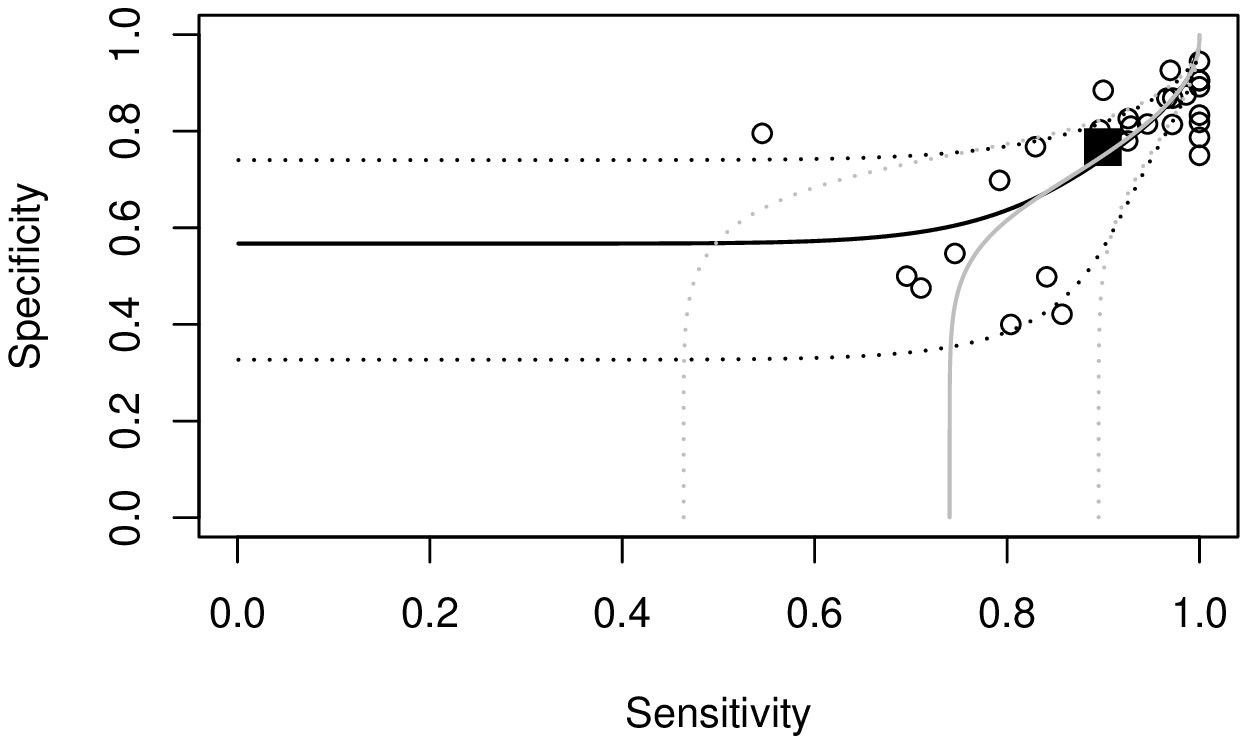} &\includegraphics[width=0.55\textwidth]{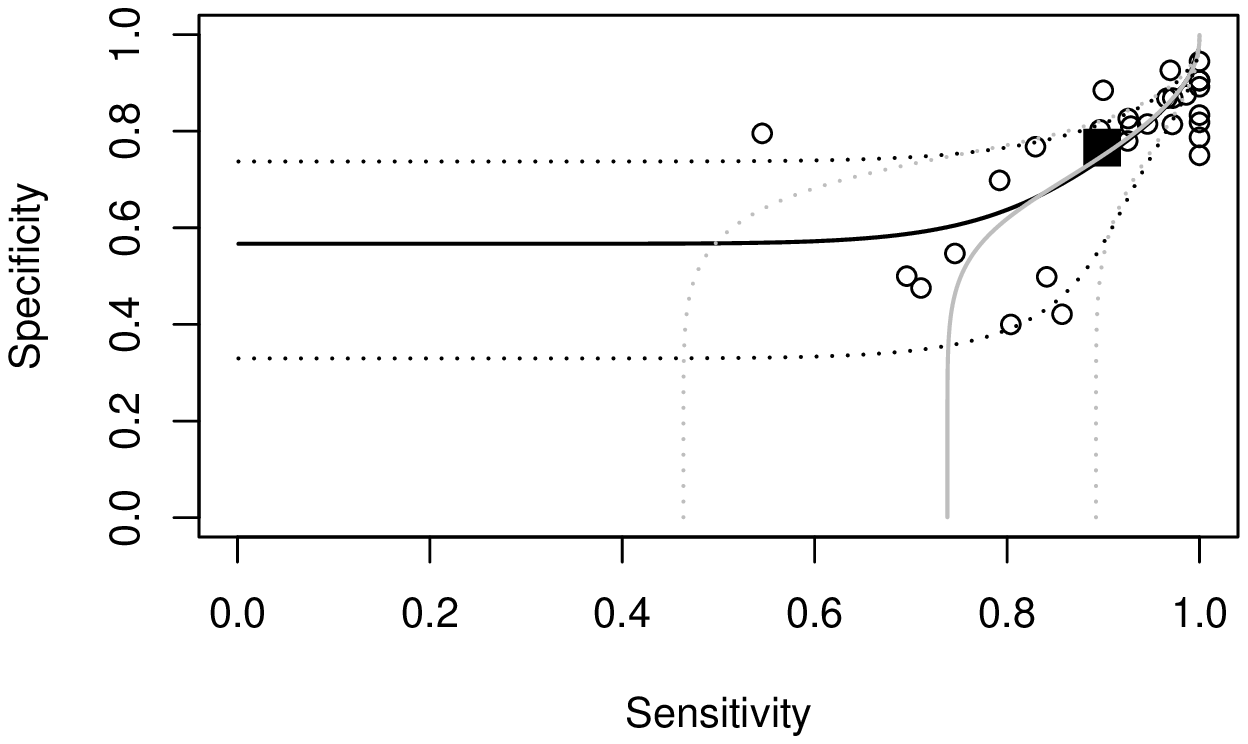}\\
\hline
\end{tabular}

\begin{flushleft}
     \begin{footnotesize}
 $\blacksquare$: summary point; $\circ$: study estimate; 
black and grey lines represent the quantile  regression curves $x_1:=\widetilde{x}_1(x_2,q)$ and $x_2:=\widetilde{x}_2(x_1,q)$, respectively; for $q=0.5$ solid lines and for $q\in\{0.01,0.99\}$ dotted lines (confidence region);    $^\P$: The non-evaluable outcomes are excluded; $^\S$: The non-evaluable positives and negatives are included as false negatives and positives, respectively.
    \end{footnotesize}
    \end{flushleft}

\begin{footnotesize}

\end{footnotesize}

\end{center}
\end{figure}
\end{landscape}

Figure \ref{SROCs} demonstrates the SROC curves  with a confidence  region and summary operating points (a pair of the model-based sensitivity  and specificity; shown by the black square) from the best fitted multinomial quadrivariate D-vine copula mixed model, the best fitted trivariate vine copula mixed model and the best fitted bivariate copula mixed models  along with the study estimates (shown by the circles in Figure  \ref{SROCs}).  
For the upper panel graphs the sensitivity and specificity at study $i$  (point estimates) have been calculated with the typical definitions of sensitivity and specificity, viz. 
\vspace{-0.2cm}
$$\frac{y_{i11}}{y_{i01}+y_{i11}} \quad \mbox{and} \quad  \frac{y_{i00}}{y_{i00}+y_{i10}},$$ 
\vspace{-0.2cm}
\noindent respectively,  as only patients with positive or negative results are considered, while for the lower panel graphs the sensitivity and specificity at study $i$  have been calculated with the definitions of sensitivity and specificity in Simel et al. \citep{Simel-etal-1987}, viz. 
\vspace{-0.2cm}
$$\frac{y_{i11}}{y_{i+1}}\quad \mbox{and} \quad \frac{y_{i00}}{y_{i+0}},$$
\vspace{-0.2cm} 
\noindent respectively, since the number  of non-evaluable positives $y_{i21}$ contributes to the diseased population and the number  of non-evaluable negatives $y_{i20}$ contributes to the non-diseased population.

\section{\label{discussion}Discussion}

Motivated by the existence of non-evaluable results in diagnostic test accuracy studies, this paper  proposed a multinomial quadrivariate D-vine copula mixed model for meta-analysis of diagnostic test accuracy studies accounting for non-evaluable subjects. Our general statistical model allows for selection of pair-copulas independently among a variety of parametric copula families, i.e. there are no constraints in the choices of bivariate parametric families of copulas and can also operate on the original scale of sensitivity and specificity. 

For the random effects, we have used a quadrivariate D-vine copula distribution or a truncated at level 1 quadrivariate D-vine copula (conditional independence), which allows flexible (tail) dependence \citep{joeetal10}. We have  proposed a numerically stable ML estimation technique based on Gauss-Legendre quadrature; the crucial step is to convert from independent to dependent quadrature points that follow a quadrivariate D-vine distribution. 

In an era of evidence-based medicine, decision makers need high-quality procedures such as the one developed in this article to support decisions about whether or not to use a diagnostic test in a specific clinical situation.
The multinomial quadrivariate vine-copula mixed model  
is not an ad-hoc \cite{Schuetz-etal-2012} but rather a sophisticated approach that  utilizes all the available data in decision making and  can satisfy the intention-to-diagnose principle.    Using an intention to diagnose principle, i.e. a conservative approach,  ensures that both the 
sensitivity and specificity are not overestimated. Hence, 
it formally enables decision makers  to be more cautious in 
solely relying 
to the  overly optimistic meta-analytic estimates of sensitivity and specificity derived from  the extended trivariate vine copula mixed model that indirectly accounts for the non-evaluable outcomes.

\section*{Software}
{\tt R} functions to implement   and simulate from the multinomial quadrivariate D-vine copula mixed model for meta-analysis of diagnostic tests with non-evaluable subjects are  part of the  {\tt  R} package {\tt  CopulaREMADA} \citep{Nikoloulopoulos-2018-CopulaREMADA}. The data and code  used in Section \ref{application} are given as data and code examples in the package, respectively.

\section*{Acknowledgements}
We would like to thank the referees for insightful  comments leading to an improved presentation. The simulations presented in this paper were carried out on the High Performance Computing Cluster supported by the Research and Specialist Computing Support service at the University of East Anglia.


\begin{thebibliography}{10}
\providecommand{\url}[1]{\texttt{#1}}
\providecommand{\urlprefix}{URL }
\expandafter\ifx\csname urlstyle\endcsname\relax
  \providecommand{\doi}[1]{DOI:\discretionary{}{}{}#1}\else
  \providecommand{\doi}{DOI:\discretionary{}{}{}\begingroup
  \urlstyle{rm}\Url}\fi
\providecommand{\eprint}[2][]{\url{#2}}

\itemsep=4pt

\bibitem{Begg-etal-1986}
Begg CB, Greenes RA and Iglewicz B.
\newblock The influence of uninterpretability on the assessment of diagnostic
  tests.
\newblock \emph{Journal of Chronic Diseases} 1986; 39(8): 575--584.

\bibitem{Schuetz-etal-2012}
Schuetz GM, Schlattmann P and Dewey M.
\newblock Use of 3$\times$2 tables with an intention to diagnose approach to
  assess clinical performance of diagnostic tests: Meta-analytical evaluation
  of coronary {C}{T} angiography studies.
\newblock \emph{BMJ (Online)} 2012; 345: e6717.

\bibitem{JacksonRileyWhite2011}
Jackson D, Riley R and White IR.
\newblock Multivariate meta-analysis: Potential and promise.
\newblock \emph{Statistics in Medicine} 2011; 30(20): 2481--2498.

\bibitem{MavridisSalanti13}
Mavridis D and Salanti G.
\newblock A practical introduction to multivariate meta-analysis.
\newblock \emph{Statistical Methods in Medical Research} 2013; 22(2): 133--158.

\bibitem{ma-etal-2014}
Ma X, Suri MFK and Chu H.
\newblock A trivariate meta-analysis of diagnostic studies accounting for
  prevalence and non-evaluable subjects: Re-evaluation of the meta-analysis of
  coronary {C}{T} angiography studies.
\newblock \emph{BMC Medical Research Methodology} 2014; 14: 128.

\bibitem{Nikoloulopoulos-2018-3dmeta-NE}
Nikoloulopoulos AK.
\newblock An extended trivariate vine copula mixed model for meta-analysis of
  diagnostic studies in the presence of non-evaluable outcomes.
\newblock \emph{The International Journal of Biostatistics} 2020.  \href{http://dx.doi.org/10.1515/ijb-2019-0107}{DOI: 10.1515/ijb-2019-0107}.

\bibitem{Chu&Cole2006}
Chu H and Cole SR.
\newblock Bivariate meta-analysis of sensitivity and specificity with sparse
  data: a generalized linear mixed model approach.
\newblock \emph{Journal of Clinical Epidemiology} 2006; 59(12): 1331--1332.

\bibitem{Nikoloulopoulos2015c}
Nikoloulopoulos AK.
\newblock A vine copula mixed effect model for trivariate meta-analysis of
  diagnostic test accuracy studies accounting for disease prevalence.
\newblock \emph{Statistical Methods in Medical Research} 2017; 26(5):
  2270--2286.

\bibitem{chu-etal-2009}
Chu H, Nie L, Cole SR et~al.
\newblock Meta-analysis of diagnostic accuracy studies accounting for disease
  prevalence: Alternative parameterizations and model selection.
\newblock \emph{Statistics in Medicine} 2009; 28(18): 2384--2399.

\bibitem{Nikoloulopoulos2015b}
Nikoloulopoulos AK.
\newblock A mixed effect model for bivariate meta-analysis of diagnostic test
  accuracy studies using a copula representation of the random effects
  distribution.
\newblock \emph{Statistics in Medicine} 2015; 34: 3842--3865.

\bibitem{Bedford&Cooke02}
Bedford T and Cooke RM.
\newblock Vines - a new graphical model for dependent random variables.
\newblock \emph{Annals of Statistics} 2002; 30: 1031--1068.

\bibitem{aasetal09}
Aas K, Czado C, Frigessi A et~al.
\newblock Pair-copula constructions of multiple dependence.
\newblock \emph{Insurance: {M}athematics \& {E}conomics} 2009; 44: 182--198.

\bibitem{nikoloulopoulos&joe&li11}
Nikoloulopoulos AK, Joe H and Li H.
\newblock Vine copulas with asymmetric tail dependence and applications to
  financial return data.
\newblock \emph{Computational Statistics \& Data Analysis} 2012; 56: 659--3673.

\bibitem{Killiches&Czado-2018}
Killiches M and Czado C.
\newblock A {D}-vine copula-based model for repeated measurements extending
  linear mixed models with homogeneous correlation structure.
\newblock \emph{Biometrics} 2018; 74(3): 997--1005.

\bibitem{nikoloulopoulos-2018-smmr}
Nikoloulopoulos AK.
\newblock A {D}-vine copula mixed model for joint meta-analysis and comparison
  of diagnostic tests.
\newblock \emph{Statistical Methods in Medical Research} 2019; 28(10-11):
  3286--3300.

\bibitem{nikoloulopoulos&joe12}
Nikoloulopoulos AK and Joe H.
\newblock Factor copula models for item response data.
\newblock \emph{Psychometrika} 2015; 80: 126--150.

\bibitem{czado-etal-12-statMod}
Czado C, Schepsmeier U and Min A.
\newblock Maximum likelihood estimation of mixed {C}-vines with application to
  exchange rates.
\newblock \emph{Statistical Modelling} 2012; 12(3): 229--255.

\bibitem{wilson2018}
Wilson KJ.
\newblock Specification of informative prior distributions for multinomial
  models using vine copulas.
\newblock \emph{Bayesian Analysis} 2018; 13(3): 749--766.

\bibitem{nash90}
Nash J.
\newblock \emph{Compact Numerical Methods for Computers: Linear Algebra and
  Function Minimisation}.
\newblock New York: Hilger, 1990.
\newblock 2nd edition.

\bibitem{Stroud&Secrest1966}
Stroud AH and Secrest D.
\newblock \emph{Gaussian Quadrature Formulas}.
\newblock Englewood Cliffs, NJ: Prentice-Hall, 1966.

\bibitem{Nikoloulopoulos-2016-SMMR}
Nikoloulopoulos AK.
\newblock Hybrid copula mixed models for combining case-control and cohort
  studies in meta-analysis of diagnostic tests.
\newblock \emph{Statistical Methods in Medical Research} 2018; 27(8):
  2540--2553.

\bibitem{Nikoloulopoulos2018-AStA}
Nikoloulopoulos AK.
\newblock On composite likelihood in bivariate meta-analysis of diagnostic test
  accuracy studies.
\newblock \emph{AStA Advances in Statistical Analysis} 2018; 102: 211--227.

\bibitem{HultLindskog02}
Hult H and Lindskog F.
\newblock {Multivariate extremes, aggregation and dependence in elliptical
  distributions}.
\newblock \emph{{Advances in Applied Probability}} {2002}; {34}: {587--608}.

\bibitem{genest87}
Genest C.
\newblock Frank's family of bivariate distributions.
\newblock \emph{Biometrika} 1987; 74(3): 549--555.

\bibitem{genest&mackay86}
Genest C and MacKay J.
\newblock The joy of copulas: bivariate distributions with uniform marginals.
\newblock \emph{The American Statistician} 1986; 40(4): 280--283.

\bibitem{paul-etal-2010}
Paul M, Riebler A, Bachmann LM et~al.
\newblock Bayesian bivariate meta-analysis of diagnostic test studies using
  integrated nested laplace approximations.
\newblock \emph{Statistics in Medicine} 2010; 29(12): 1325--1339.

\bibitem{Menke&Kowalski2016}
Menke J and Kowalski J.
\newblock Diagnostic accuracy and utility of coronary ct angiography with
  consideration of unevaluable results: A systematic review and multivariate
  bayesian random-effects meta-analysis with intention to diagnose.
\newblock \emph{European Radiology} 2016; 26(2): 451--458.

\bibitem{joeetal10}
Joe H, Li H and Nikoloulopoulos AK.
\newblock Tail dependence functions and vine copulas.
\newblock \emph{Journal of Multivariate Analysis} 2010; 101: 252--270.

\bibitem{Rucker-schumacher-2009}
R\"ucker G and Schumacher M.
\newblock Letter to the editor.
\newblock \emph{Biostatistics} 2009; 10(4): 806--807.

\bibitem{Simel-etal-1987}
Simel DL, Feussner JR, Delong ER et~al.
\newblock Intermediate, indeterminate, and uninterpretable diagnostic test
  results.
\newblock \emph{Medical Decision Making} 1987; 7(2): 107--114.

\bibitem{Nikoloulopoulos-2018-CopulaREMADA}
Nikoloulopoulos AK.
\newblock \emph{{CopulaREMADA}: {C}opula mixed models for multivariate
  meta-analysis of diagnostic test accuracy studies}, 2019.
\newblock {R} package version 1.3. URL:
  \href{http://CRAN.R-project.org/package=CopulaREMADA}{http://CRAN.R-project.org/package=CopulaREMADA}.

\end{thebibliography}

\end{document}